\documentclass[prl,aps,floats,superscriptaddress,twocolumn]{revtex4-1}

\usepackage[LGR,T1]{fontenc}
\usepackage{amstext}
\usepackage{amssymb}
\usepackage{stmaryrd}
\usepackage{graphicx}
\usepackage{subscript}
\usepackage{amsmath} 
\usepackage{latexsym}
\usepackage[english]{babel}
\usepackage{blindtext}
\usepackage{graphicx}
\usepackage{subfigure}
\usepackage{indentfirst}
\usepackage{appendix}
\usepackage{mathrsfs}
\usepackage{color}
\usepackage[colorlinks,linkcolor=blue,anchorcolor=blue,citecolor=green]{hyperref}
\usepackage{ulem}
\usepackage{float}
\usepackage{subfiles}
\allowdisplaybreaks[2]

\newcommand {\no} {\nonumber}
\newcommand{\lt} {\left}
\newcommand{\rt} {\right}

\newcommand{\nocontentsline}[3]{}
\newcommand{\tocless}[2]{\bgroup\let\addcontentsline=\nocontentsline#1{#2}\egroup} 

\ProvideTextCommand{\~}{LGR}[1]{\char126#1}

\makeatother

\begin{document}

	\title{Tuning Superinductors by Quantum Coherence Effects for Enhancing Quantum Computing}
	\author{Bo Fan}
	\email{bo.fan@sjtu.edu.cn}
	\affiliation{Shanghai Center for Complex Physics, 
		School of Physics and Astronomy, Shanghai Jiao Tong
		University, Shanghai 200240, China}
	\author{Abhisek Samanta}\email{abhiseks@campus.technion.ac.il}
	\affiliation{ Physics Department, Technion, Haifa 32000, Israel}
	\author{Antonio M. Garc\'ia-Garc\'ia}
	\email{amgg@sjtu.edu.cn}
	\affiliation{Shanghai Center for Complex Physics, 
		School of Physics and Astronomy, Shanghai Jiao Tong
		University, Shanghai 200240, China}
	\begin{abstract}
			Research on spatially inhomogeneous weakly-coupled superconductors has recently received a boost of interest because of the experimental observation of a dramatic enhancement of the kinetic inductance with relatively low losses. Here, we study the kinetic inductance and the quality factor of a strongly-disordered weakly-coupled superconducting thin film. We employ a gauge-invariant random-phase approximation capable of describing collective excitations and other fluctuations. In line with the experimental findings, we have found that in the range of frequencies of interest, and for sufficiently low temperatures, an exponential increase of the kinetic inductance with disorder coexists with a still large quality factor $\sim 10^4$.  More interestingly, on the metallic side of the superconductor-insulator transition, we have identified a range of frequencies and temperatures $T \sim 0.1T_c$ where quantum coherence effects induce a broad statistical distribution of the quality factor with an average value that increases with disorder. We expect these findings to further stimulate experimental research on the design and optimization of superinductors for a better performance and miniaturization of quantum devices such as qubit circuits and microwave detectors.
		\end{abstract}
	
	\maketitle
A microwave resonator that is both low-loss and small \cite{you2002,gu2017microwave,kjaergaard2020} is an important element in circuits designed for quantum computation or photon detection. For instance, a high-inductance element \cite{peltonen2013} suppresses charge fluctuations and slows down electromagnetic waves which increases coherence times and facilitates the circuit miniaturization.
Devices such as the fluxonium qubit \cite{manuchar2009}
have a kinetic inductance $L_{k/\square}$ orders of magnitude larger than the geometrical one related to its shape and it is further enhanced by increasing disorder.   
For that reason, there have been an explosion of interest \cite{manuchar2009,annunziata2010,niepce2019,zhang2019,rotzinger2016,valenti2019,lukas2018,maleeva2018,lukas2019,deutscher2020,winkel2020,fang2020,dupre2017,de2018,coumou2013} in disordered superconductors as high-inductance devices, commonly called superinductors. Promising results \cite{rotzinger2016,valenti2019,lukas2018,maleeva2018,lukas2019,deutscher2020,winkel2020,fang2020} have been reported for  
nano-wires of NbN$_x$, tin, granular Al \cite{Deutscher1973,Deutscher1973a,Abeles1977,Abeles1966,meservey1969} and nano-wires of Al based materials \cite{annunziata2010,niepce2019,zhang2019,coumou2013}.

These materials have also drawbacks. The presence of sub-gap collective excitations \cite{anderson1958sc,anderson1963,mattis1958}, specially the Goldstone mode \cite{goldstone1961} related to phase fluctuations, in highly disordered superconductors 
\cite{seibold2017,cea2014,cea2015,abhisek2020,bofan2021a} may shorten coherence times in superconducting circuits. 
Another potential problem is the existence of strong losses \cite{martinis2005,gao2008} in amorphous dielectric due to two level systems. Although this source of decoherence can to some extent be controlled, for instance by reducing the device size \cite{martinis2005,gao2008}, it can reduce the coherence time of the device. However, the current experimental evidence \cite{lukas2018,deutscher2020} is that this effect is not dominant.

Quantum coherence effects play an important role in weakly-coupled disordered superconductors. The order parameter becomes highly inhomogeneous  \cite{Mayoh2015,bofan2020,bofan2020a,Burmistrov2012,Mayoh2014,Mayoh2014a,Burmistrov2015,tezuka2010,Feigelman2007} with a multifractal-like spatial structure close to the insulating transition leading to the enhancement of superconductivity \cite{Mayoh2015}. Level degeneracies \cite{Parmenter1968a,Shanenko2007,garcia2008bardeen,Garcia-Garcia2011} 
cause similar effects in granular materials \cite{Mayoh2014a,Mayoh2014} and single superconducting grains.
These theoretical findings have been largely confirmed experimentally \cite{brun2014,xue2019,verdu2018,Deutscher1973,pracht2016,pracht2017,Bose2010,li2015,Brihuega2011}.

The above results call for a detailed theoretical study of quantum coherence effects in the  
kinetic inductance and quality factor of disordered superconducting thin films. 

In this paper, we address this problem in a two dimensional weakly-coupled disordered superconductor close to the insulating transition but still on the metallic side. We compute these observables by the mean-field limit of the attractive Hubbard model in the weak-coupling limit, namely, the self-consistent Bogoliubov de-Gennes (BdG) formalism \cite{DeGennes1964,DeGennes1966}. Corrections to the BdG formalism are obtained by a random phase approximation leading to gauge-invariant results. Diagrammatically, these deviations correspond to vertex corrections to the mean-field bubble-diagrams representing the current-current correlation function. Gauge invariance is necessary in order to reproduce sub-gap collective excitations such as the Goldstone mode \cite{goldstone1961} that may increase losses dramatically and therefore prevent the use of disordered superconductors as high-inductance devices. We shall see that, in contrast with previous results \cite{ioffe2018} on the insulating side of the transition, collective excitations do not occur in the region of interest. Moreover, in agreement with experiments, the kinetic inductance increases with disorder without a drastic reduction of the quality factor. 
\medskip
\noindent

{\it Kinetic inductance and Quality factor of a superinductor device.\textemdash}
The reactance is defined as $X = 2\pi f L_{k/\square}$, so a high performance superinductor requires a large kinetic inductance $L_{k/\square}$. According to the Mattis-Bardeen theory \cite{mattis1958}, the kinetic inductance of a superconductor is $L_{k/\square} \sim {\hbar R_\square \over \pi \Delta}$, where $R_\square$ is the sheet resistance of the material and $\Delta$ is the amplitude of the superconducting order parameter. Therefore, $L_{k/\square}$ is enhanced by increasing the sheet resistance, which effectively is equivalent to an increase of the disorder strength. However, as mentioned earlier, disorder can also lead to a stronger dissipation at micro-wave frequencies due to collective modes and other fluctuations \cite{cea2014,abhisek2020,bofan2021a} that induce sub-gap structure in the conductivity near the insulating transition. 
Therefore, it is important that the enhancement of the kinetic inductance by disorder is not accompanied by these dissipative effects. 

\begin{figure}
 \begin{center}
				\subfigure[]{\label{fig.comp_a} 
						\includegraphics[width=4.2cm]{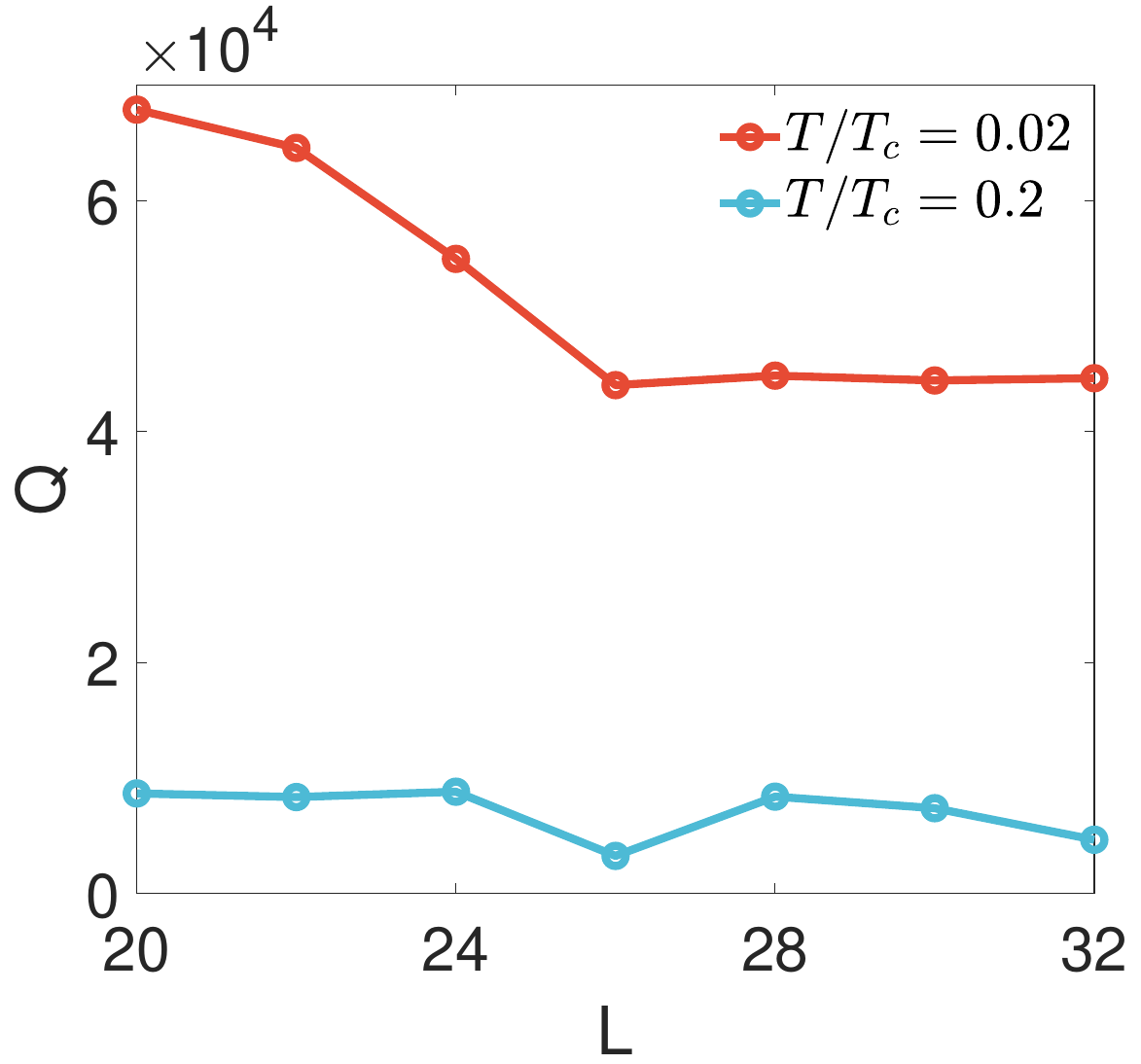}}
				\subfigure[]{\label{fig.comp_c} 
						\includegraphics[width=4.2cm]{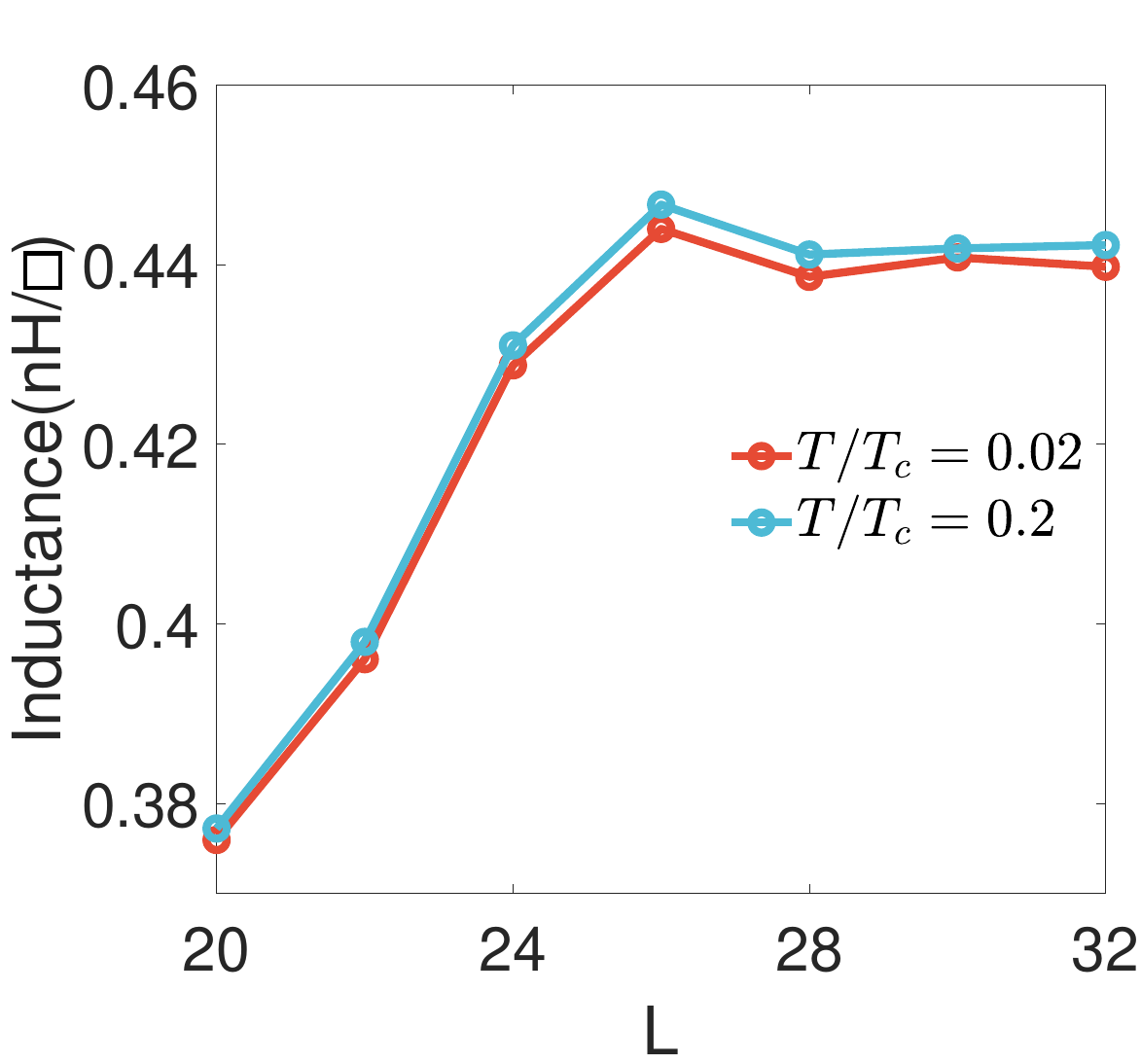}}
		\caption{The quality factor $Q$ \subref{fig.comp_a} and the kinetic inductance $L_{k/\square}$ \subref{fig.comp_c} as a function of system size $L$ for $U = -1, \langle n \rangle = 0.875, f = 0.028\Delta_0$, and $V = 1.5$, close to the critical disorder, which is the main focus of the paper. No substantial size dependence is observed for $L \geq 26$, so we fix the size $L = 28$ in our study. For stronger disorder or coupling, finite size effects will be even smaller.} \label{Fig:comp}
	\end{center}
\end{figure}
Dissipation effects are described by the quality factor $Q = {f \over {\Delta f}}$, where $f$ is the circuit operating frequency (measured in Hertz). $Q$ describes the frequency resolution $\Delta f$ which is closely related to the strength of dissipative effects leading to electromagnetic absorption and therefore to circuit losses.
The total quality factor of the circuit has two components, ${1\over Q} = {1\over Q_i} + {1\over Q_{ext}} $ \cite{hafner2014surface} where $Q_i$ is the internal quality factor computed in the paper due to losses in the superinductor, and $Q_{ext}$ is the external quality factor related to any other losses. $Q_{ext}$ is typically much higher than $Q_i$ in the experimental settings that we are interested in, so $Q \approx Q_i$.

 Two superinductor geometries are especially relevant for applications:  superconducting micro-strips lines for quantum computing and cavity resonators for phonon detection. We shall see that the main findings of the paper are applicable to both settings though $Q$ in each case is different by an overall disorder and temperature independent prefactor. From now on, we focus on the micro-strip setup, the cavity resonator is discussed in the Supplemental information \cite{suppl}.     
  The quality factor of the micro-strip is \cite{yassin1995electromagnetic,devisser2014,cheng1989field} $Q = {\alpha\over 2\beta}$, where $\alpha$ and $\beta$ are the real and imaginary parts of the propagation constant $\gamma$. 
 For dissipative media, \cite{yassin1995electromagnetic,cheng1989field} $\gamma = \sqrt{i\omega \mu(\sigma+i\omega\epsilon)} = \alpha + i\beta$, where $\epsilon$ and $\mu$ are the permittivity and permeability of the media, and $\sigma = \sigma_1 - i\sigma_2$, where $\sigma_1$ and $\sigma_2$ are the real and imaginary parts of the conductivity $\sigma$. 
Since $\sigma_2 \gg \sigma_1$, and the device operates at microwave frequencies \cite{devisser2014,yassin1995electromagnetic}, $Q \approx \sigma_2 /\sigma_1$. 
%
%
%
Likewise, the kinetic inductance is $L_{k/\square} = {1\over 2\pi f \sigma_2}$.

\medskip
\noindent
{\it Theoretical formalism.\textemdash} In order to compute $L_{k/\square}$ and $Q$,
we model the resonator as a disordered superconducting thin film with $N = L\times L$ sites described by the BdG equations \cite{DeGennes1964, DeGennes1966, Ghosal2001, ghosal1998},
\begin{equation}
	\left(\begin{matrix}
		\hat{K} 		& \hat{\Delta}  \\
		\hat{\Delta}^* 	& -\hat{K}^* 	\\
	\end{matrix}\right)
	\left(\begin{matrix}
		u_m(i)   \\
		v_m(i) \\
	\end{matrix}\right)
	= E_m
	\left(\begin{matrix}
		u_m(i)   \\
		v_m(i) \\
	\end{matrix}\right)
	\label{eq.bdg}
\end{equation} 
where $\hat{K}u_m(i)=-t\sum_{\delta}u_m(i+\delta)+(V_i-\mu_i)u_m(i)$, $V_i$ is a random potential $V_i\in [-V, V]$, $\mu_i = \mu + |U|n(i)/2$ is the chemical potential that incorporates the site-dependent Hartree shift, $U$ is the net attractive electronic coupling, and $\delta$ is restricted to nearest neighbors of site $i$. The BdG equations are completed by the self-consistent conditions for the site-dependent order parameter amplitude $\Delta(i) = |U|\sum_{m}u_m(i)v_m^*(i)(1-2f(E_m,T))$ and the density $n(i) = 2\sum_{m}[ |v_m(i)|^2(1-f(E_m,T)) + |u_m(i)|^2f(E_m,T) ]$ where $f(E_m,T) = \frac{1}{e^{E_m/T}+1}$ is the Fermi-Dirac distribution at temperature $T$. The average electron density is $\langle n \rangle =\sum_{i}n(i)/N=0.875$ with $N$ the total number of sites of the square lattice. This choice, previously used in \cite{ghosal1998,Ghosal2001,seibold2017,datta2021}, leads to \cite{huscroft1998} suppression of charge-density wave correlations while keeping the order parameter as large as possible. All these quantities are in units of the hopping energy $t$.
Based on the solutions of these equations, we compute current-current correlation functions beyond the mean-field BdG limit within a gauge invariant random phase approximation which amounts to considering vertex corrections to the bare bubble diagrams representing the susceptibility. Finally, we compute the complex conductivity from which, as mentioned earlier, is straightforward to find $Q(T,V,U)$ and $L_{k/\square}(T,V,U)$ [see Supplemental information \cite{suppl} for technical details]. Our study is focused on the metallic side of the superconductor-insulator transition where we will also compute the distribution of probability of $Q$ to assess the role of multifractality and to have an estimate of sample to sample fluctuations which are important in experiments. 
The choice of parameters in our calculation is motivated by the following considerations: 
a superconductor resonator typically works in the microwave region $1-20$ GHz \cite{gu2017microwave,grunhaupt2019granular,niepce2019,krantz2019quantum,valenti2019,sergeev2002ultrasensitive,samkharadze2016high}, which in units of energy corresponds to $4.0\times10^{-3} - 8.0\times10^{-2} $ meV. 
Therefore, we set the hopping energy to $t = 10$ meV so that the operating frequencies of the device lie in the range $1-20$ GHz. We will focus on two typical frequencies $f \sim 0.001t = 0.028\Delta_0$ and $0.004t = 0.112\Delta_0$ corresponding to $2.5$ GHz and $10$ GHz respectively. The lattice constant is set to $a = 0.35$ nm and the permittivity $\epsilon = 9 \epsilon_0$. We note that all the calculation is carried in SI units, which can compare with experiments directly. In order to have a larger $Q$, temperatures must be much lower than the critical one i.e. $T\ll T_c$, but still accessible experimentally $T \sim 0.01-0.15 T_c$. We model weakly coupled superconductors by a coupling constant $U=-1$, in units of $t$, 
because this is the smallest coupling for which the finite size effects in our calculation are negligible. For these parameters, we obtain for $V=0$ an order parameter $\Delta_0 = 0.0357t = 0.357$ meV and a critical temperature $T_c \sim 0.02t \sim 2.3$ K.
\begin{figure}
\begin{center}
			\subfigure[]{\label{fig.Q_a} 
					\includegraphics[width=4.2cm]{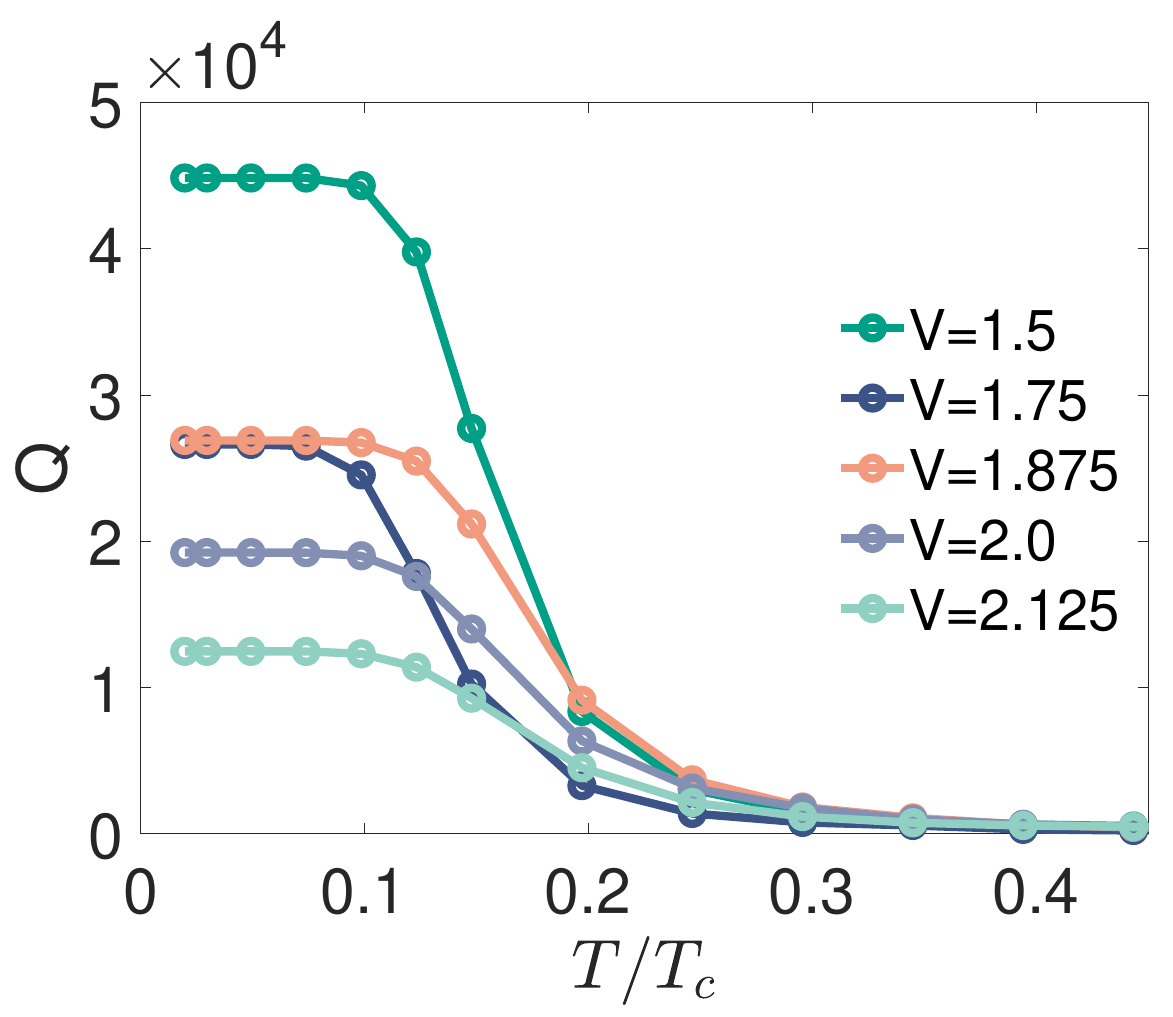}}
			\subfigure[]{\label{fig.Q_b} 
					\includegraphics[width=4.2cm]{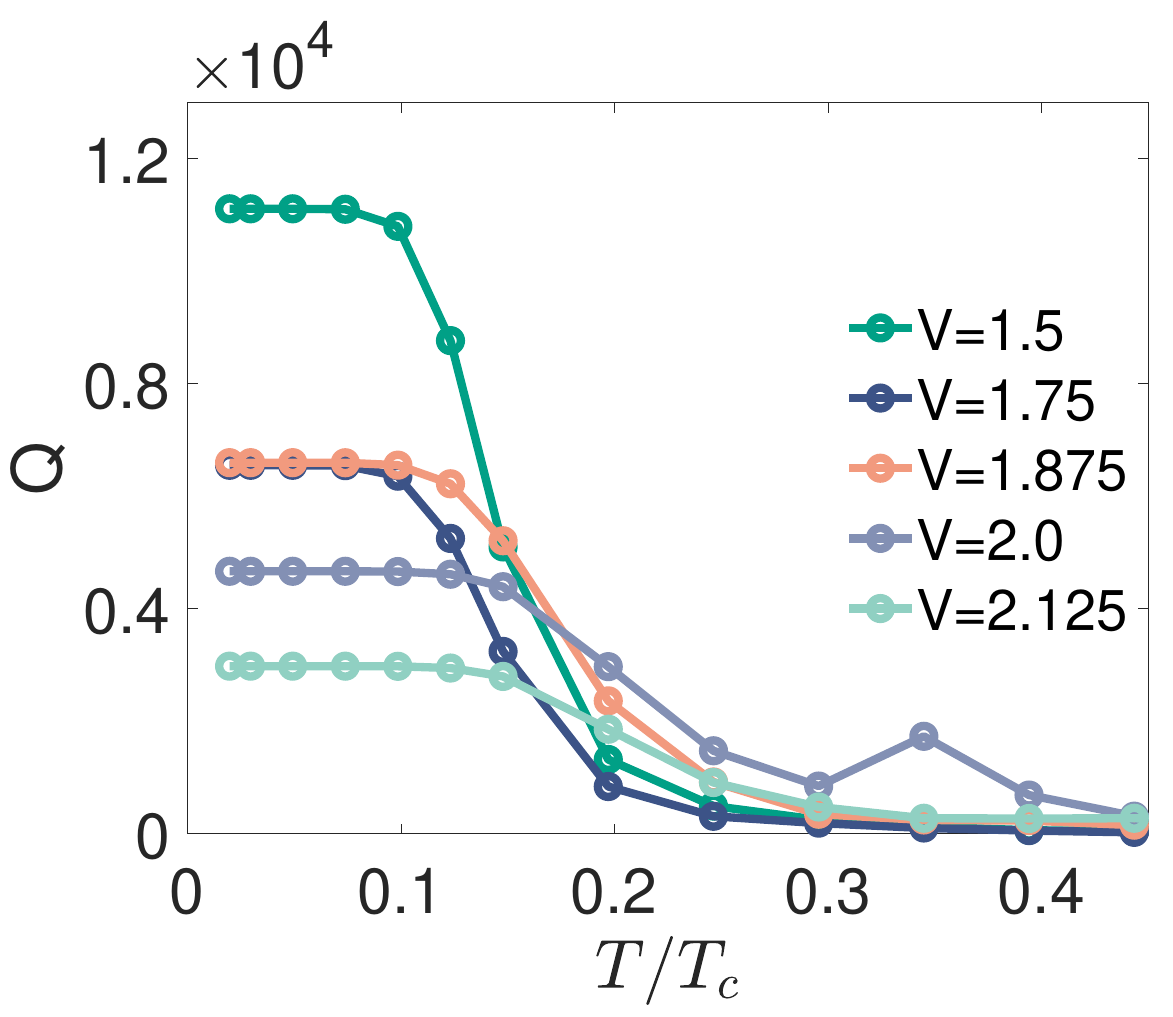}}
\caption{Quality factor $Q$ as a function of temperature for a lattice of size $L = 28$ and different disorder strengths $V$, $U = -1$ and $\langle n \rangle = 0.875$. $Q$ is computed using $70$ disorder realization for each $V$, except for $V=2.125$ where we employ only $20$. The frequency is, \subref{fig.Q_a} $f =  0.028\Delta_0 \sim 2.5$ GHz \subref{fig.Q_b} $f = 0.112\Delta_0 \sim 10$ GHz.   }\label{Fig:Q_U1}
\end{center}
\end{figure}

\medskip
\noindent
{\it Finite-size effects.\textemdash}
We start our analysis of $L_{k/\square}$ and $Q$ by showing that finite size effects are not important. 
In Fig.~\ref{Fig:comp}, we depict $Q$ for different system sizes $L$ and for two different temperatures. We do not observe any substantial size dependence for $L\geq 26$. 
Finite size effects in $Q$ are much stronger for small sizes because quantum fluctuations increase as the system size decreases. Finite temperature effects, as observed in Fig.~\ref{Fig:comp} suppress quantum fluctuations. The increase (decrease) of $L_{k/\square}$ ($Q$) with system size is due to the fact that $L_{k/\square}$ is inversely proportional to $\sigma_2$ while $Q$ is directly proportional to it.   
For a stronger coupling constant $|U|>1$, finite size effects are smaller because the coherence length is shorter. Therefore, from now on we set $L = 28$. 

\medskip
\noindent
{\it Temperature, disorder and frequency dependence of the Quality factor.\textemdash}
At sufficiently high temperatures, $Q$ decreases sharply due to the increasing number of thermal quasiparticles \cite{sergeev2002ultrasensitive}. Therefore,  this region is not of interest in our analysis. 
In the superconducting region, ${1\over Q}$ is proportional to the remaining number of thermal quasiparticles, which will be reduced greatly by decreasing temperature~\cite{valenti2019,sergeev2002ultrasensitive}. As a consequence, $Q$ will increase substantially with decreasing temperature. However, as Fig.~\ref{Fig:Q_U1} shows, this increase levels off for sufficiently low $T \leq 0.1T_c$ which is roughly consistent with experimental results \cite{samkharadze2016high}. 
The origin of this saturation of $Q$ is due to quantum fluctuations \cite{cea2014,abhisek2020}, captured in our analysis by the employed gauge-invariant random phase approximation, that smear out the real conductivity so that it is finite even below the spectral gap. 
Since $Q$ is a decreasing function of temperature, the optimal operation of the device will always occur at the lowest temperature that can be reached experimentally though large values of $Q$ can still be observed for $T \sim 0.1T_c$. We note that we are assuming a thermal distribution of quasiparticles but there is substantial evidence that in the low temperature limit, the distribution may be non-thermal \cite{chang1977,wolter1981}. However, the quality factor measured experimentally \cite{devisser2014} is still quite large $Q > 10^4$, so we do not expect that our main results are altered qualitatively by considering a non-thermal distribution.

Regarding the dependence on disorder, for sufficiently low temperatures, $Q$ decreases sharply as disorder is increased, see Fig.~\ref{Fig:Q_U1}. This is expected as both quantum fluctuations and spatial inhomogeneities facilitate the absorption of the incoming electromagnetic radiation and therefore the decrease of $Q$. 
Despite of this decrease, $Q$ is still large $\sim 10^4$ for $f \sim 2.5$ GHz and $\sim 5\times 10^3$ for $f \sim 10$ GHz, in the sub-gap frequency region of interest provided that the disorder strength is not too strong $V \lesssim 2$ where $V$ is expressed in units of $t$. The $V > 2$ insulating region is of no interest for applications.
\begin{figure}
	\begin{center}
		\subfigure[]{\label{fig.Q_w1} 
				\includegraphics[width=4.2cm]{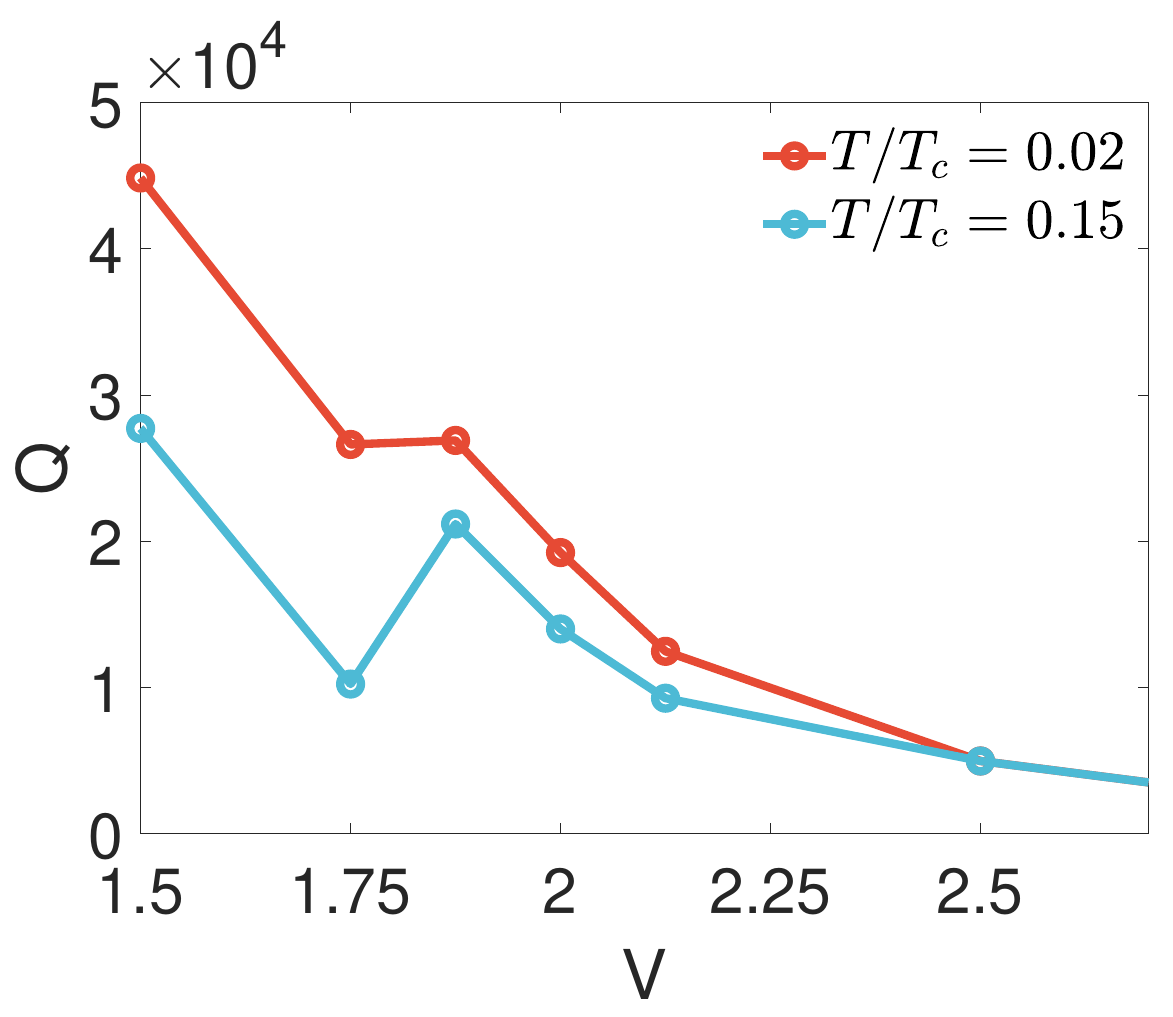}}
		\subfigure[]{\label{fig.Q_w4} 
				\includegraphics[width=4.2cm]{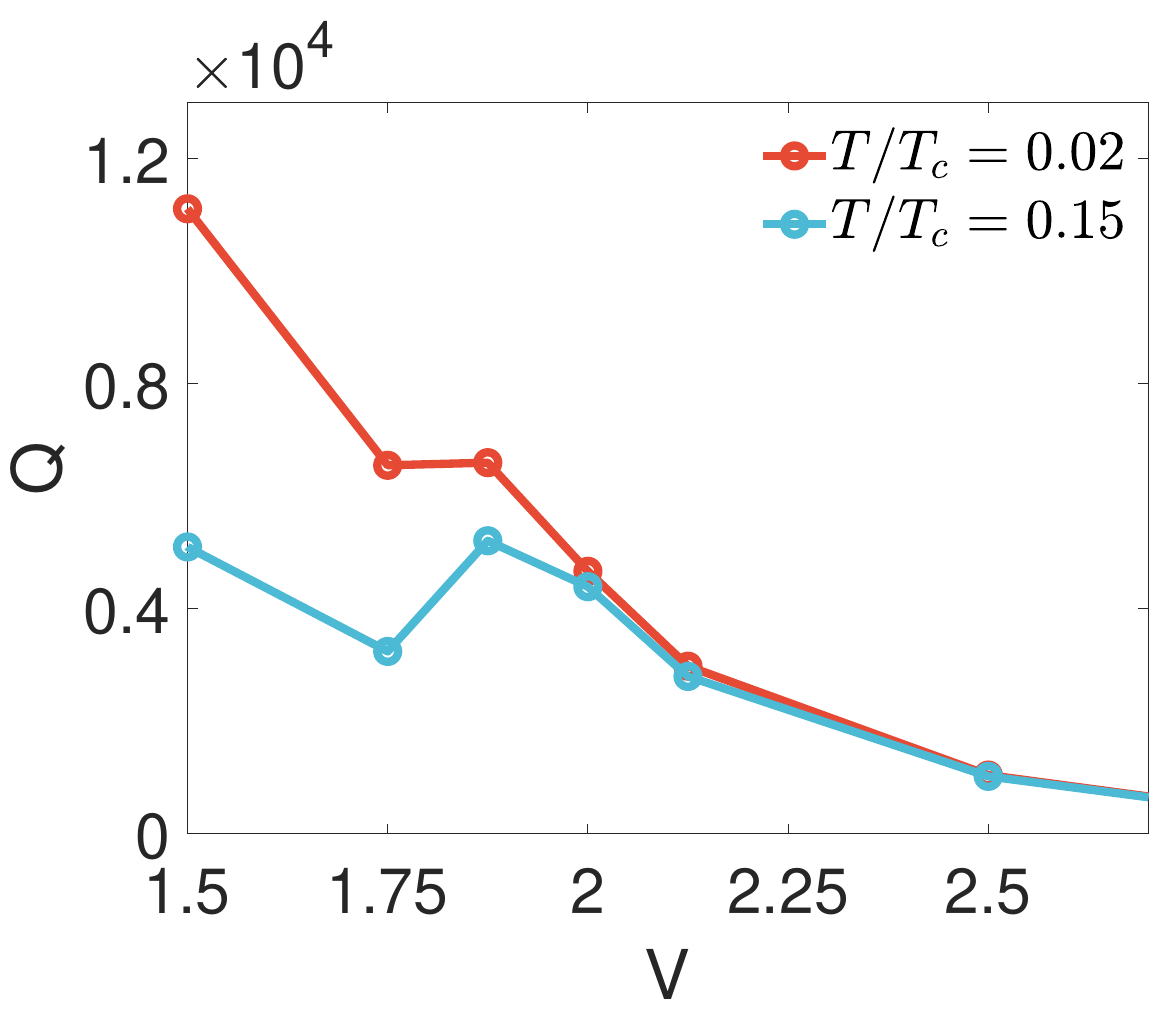}}
\caption{The quality factor $Q$ as a function of disorder $V$ for $T/T_c = 0.02$ and $0.15$ at the experimental frequencies \subref{fig.Q_w1} $f \sim 2.5$ GHz and \subref{fig.Q_w4} 
$f \sim 10$ GHz. $Q$ increases with disorder in the critical region close to the transition, $V\simeq 1.9$. This non-monotonicity of $Q$ occurs for others $T\gtrsim 0.1T_c$ and $f$, but only for $V\simeq 1.9$. See Fig.~\ref{Fig:Q_U1} for the value of the rest of parameters.}
\label{Fig:Q_dV}
\end{center}
\end{figure}
The disorder dependence of $Q$, for small but finite $T \gtrsim 0.1T_c$, shows an unexpected non-monotonous behavior in the critical region $V\sim1.9$, see Fig.~\ref{Fig:Q_U1}. In order to fully confirm it, in Fig.~\ref{Fig:Q_dV}, we depict $Q$ in linear scale as a function of $V$. The non-monotonicity is clearly observed in a relatively small window of temperatures, and only in the critical region $V \simeq 1.9$, but its effect is rather strong. Below, we provide a tentative explanation of this property which is one of the main results of the paper. The dependence of $Q$ on frequency is monotonous, it decreases rapidly with increasing frequency. The optimal setting is therefore the smallest frequency that can be accessible to experiments \cite{suppl}.

\medskip
\noindent
{\it Kinetic inductance.\textemdash}
We now turn to the study of $L_{k/\square}$. The analysis is simpler because it grows monotonically with $V$ and it is weakly dependent on frequency for frequencies below the so called two-particle gap, namely, the frequency required to break a Cooper pair \cite{tinkham2004}.
Since the resistance increases exponentially with disorder, we expect a similar behavior in the kinetic inductance $L_{k/\square}$. In qualitative agreement with the experimental results \cite{peruzzo2020,annunziata2010,niepce2019,zhang2019,deutscher2020}, see Fig.~\ref{Fig:L_k}, disorder enhances $L_{k/\square}$ by up to two orders of magnitude in the metallic region $V \lesssim 2$ where our formalism is applicable. 

\medskip
\noindent
{\it Critical disorder and optimal choice of parameters.\textemdash}
Since $L_{k/\square}$ increases sharply with disorder, and a large $Q$ requires $V \leq 2$, it is important to have an estimation of the critical disorder strength $V_c$ at which the insulating transition takes place. The result of an approximate percolation analysis suggests $V_c \sim 2$. This is consistent with an earlier \cite{bofan2020} estimate $V_c \sim 1.5$ in a similar system based on level statistics and the vanishing of the superfluid density. At this $V$, the spectral gap also starts to increase with disorder which is another indication \cite{ghosal1998,Ghosal2001} of the transition [see Supplemental information ~\cite{suppl} for more information]. 

Therefore, based also on the previous results, $V \lesssim 1.9$, corresponding to the strongest disorder still on the metallic side of the transition, is the optimal setting to observe an enhancement of $L_{k/\square}$ without a large decrease in $Q \ge 10^4$. 
Regarding the rest of parameters, based on the previous analysis, the optimal conditions for operation of the superinductor are weak-coupling $|U|\leq 1$, frequencies $f \sim 2.5-10$ GHz well below the two-particle spectral gap and the lowest accessible temperature. However, $T \gtrsim 0.15T_c \ll T_c$ is still close to this optimal value and we could also observe in $Q$ an intriguing non-monotonous dependence on disorder. 

With these choices, $L_{k/\square}$ increases up to two orders of magnitude with $V$, and this is not accompanied by strong dissipation since $Q \gtrsim 10^4$ even for $V\lesssim 1.9$. In order to describe quantitatively the experimental results, it would be necessary to have a precise relation between $V$ and the experimental resistivity $\rho_{dc}$ which is a difficult task. We can only say that the region $V \sim 1.9$ of main interest corresponds experimentally to the metallic side of the transition. Perturbatively, $\rho_{dc} \propto \langle V^2 \rangle$ for a non-interacting disordered metal \cite{datta2021}.
\begin{figure}  
	\begin{center}
		\subfigure[]{\label{fig.Lk} 
				\includegraphics[width=4.2cm]{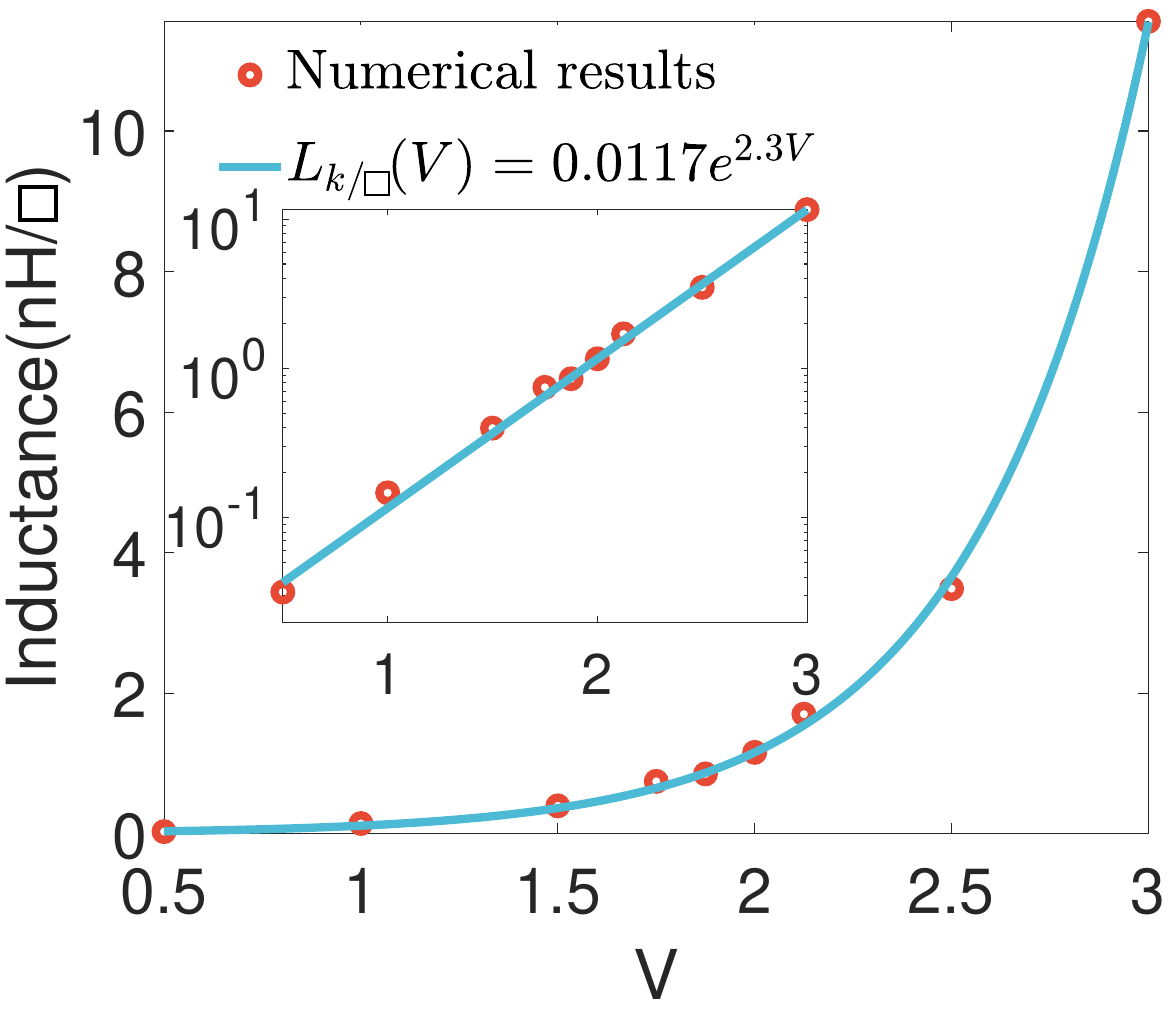}}
		\subfigure[]{\label{fig.Q_pro} 
			        \includegraphics[width=4.2cm]{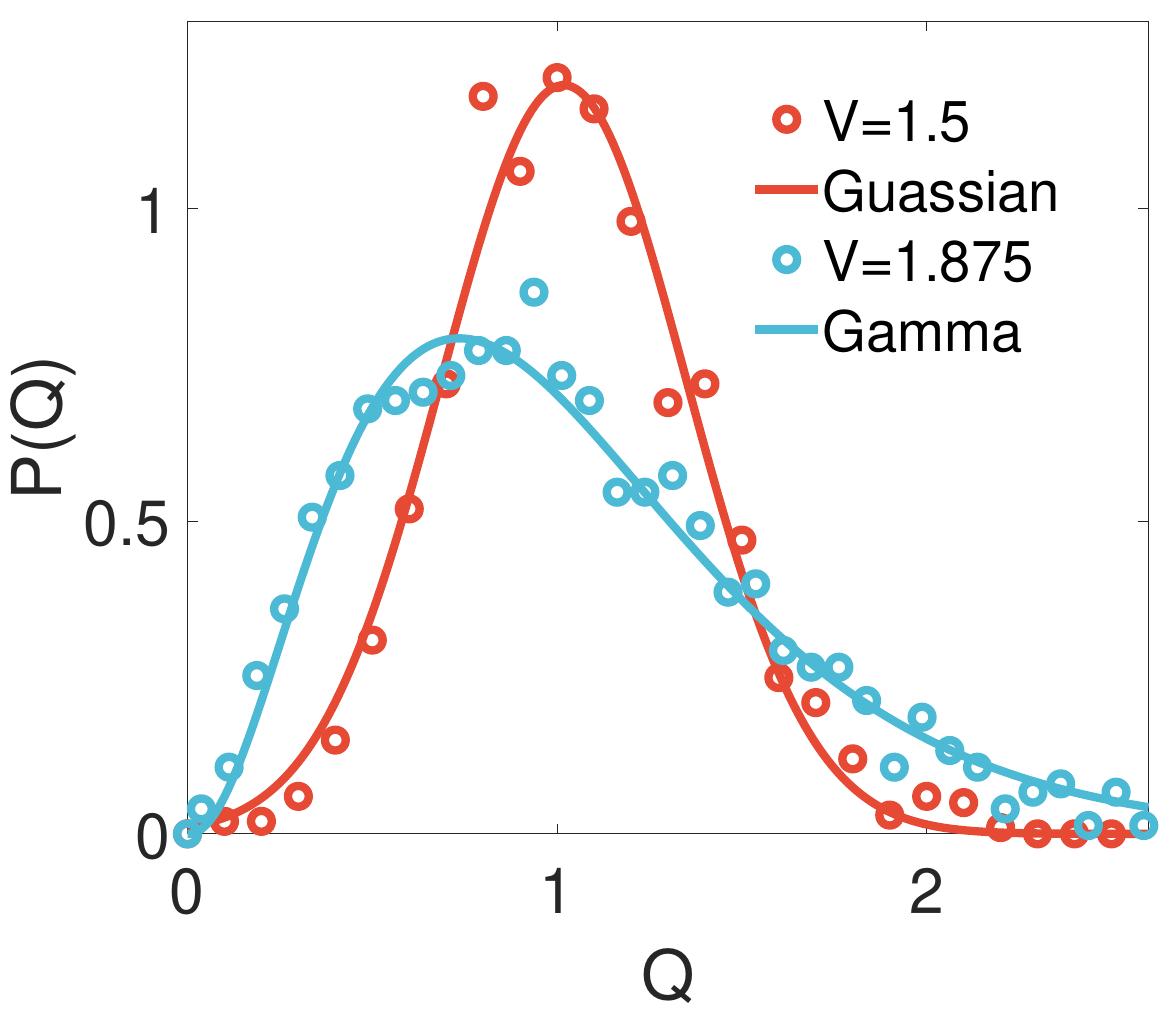}}
\caption{Left: The kinetic inductance $L_{k/\square}$ as a function of disorder $V$. The other parameters are the same as those in Fig.~\ref{Fig:Q_U1}. The numerical results fit well with an exponential increase of $L_{k/\square}$ with disorder. In the Supplemental Material~\cite{suppl} we show that in the region of interest $f \in [2.5,10]$ GHz, $T \ll T_c$, $L_{k/\square}$ is almost temperature and frequency independent. Right: Probability distribution of $Q$ from $1000$ disorder realizations, normalized by its sample averaged value for $f \sim 2.5$ GHz, and temperature $T/T_c = 0.05$. When $V=1.5$, the distribution of $Q$ fits well with a Gaussian distribution, red line. However, very close to the transition $V \approx 1.9\approx V_c$, the distribution, asymmetric and with broad tails, is similar to a Gamma distribution (blue line).}
\label{Fig:L_k}
\end{center}
\end{figure}

\medskip
\noindent
{\it Discussion and conclusions.\textemdash}
Previously, we found that close to the transition $V_c \simeq 1.9$, and for $T \gtrsim 0.1T_c$, disorder can even enhance the quality factor. We first address the origin of this counter-intuitive feature. 
In principle, by increasing disorder, the order parameter is weakened in parts of the sample which become optically active more easily and lead to the decrease of $Q$. 
However, the order parameter spatial distribution close to the transition has multifractal-like features \cite{bofan2020,bofan2020a} like a broad log-normal probability distribution \cite{mayoh2015global} which adds a twist to this argument. At zero or low temperature $T\ll T_c$, metallic regions with a large value of the order parameter amplitude coexist with regions where it is close to zero or with incipient Anderson localization effects. 

By a slight increase in disorder, it is plausible that regions with a large value of the order parameter remain largely unaltered while the rest may effectively become an Anderson insulator. Unlike metallic but not superconducting regions, those insulating regions do not absorb the incoming electromagnetic radiation. Its net effect is a reduction of optically active regions and therefore an increase of $Q$ with disorder.

As a consequence, the observed sharp drop of $Q$ with temperature occurs at higher temperatures, see Fig.~\ref{fig.Q_a}, and there exists a relatively narrow region of temperatures $T \sim 0.15T_c$, for $V \approx V_c$, where the $V$ dependence on $Q$ is non-monotonic, namely, $Q$ increases with disorder for  $V\sim V_c$.

The qualitative changes of $Q$ close to the transition $V_c \simeq  1.9$ (see Fig.~\ref{Fig:L_k}) are illustrated by computing the probability distribution obtained from $1000$ disorder realizations. Even for $V=1.5$, the distribution is close to a Gaussian. However, for $V_c \approx 1.9$, it becomes broad and asymmetric signaling large sample to sample fluctuations. This is the region where the order parameter spatial distribution has multifractal-like features so we believe that the anomalous enhancement of $Q$ is related to this feature.
We stress that the employed theoretical formalism is gauge invariant and therefore it is capable to account for collective excitations \cite{nambu1960,anderson1963} that lower dramatically the quality factor of the device. We have found \cite{bofan2021a} that indeed collective excitations may occur in weakly-coupled  disordered superconductors. However, it is necessary to have a disorder larger than $V_c \simeq 1.9$ for the collective excitations to be observed for the frequencies $2.5-10$ GHz of interest.
Therefore, collective excitations do not seem to play any role in the superinductor optimal region of operation.

The coupling we employ $U=-1$ and $\langle n \rangle = 0.875$
is too strong to describe granular Al \cite{deutscher2020} experiments. However, we expect similar \cite{waldram1976} physics is at play though the interplay of collective modes and granularity require a separate study. 
Note that we neglect Coulomb interactions. It has been recently argued \cite{glezer2021,deutscher2020} that Coulomb interactions may play an important role in granular Al close to the insulating transition.  
However, on the metallic region of interest, charging effects are screened and therefore do not play an important role.

In conclusion, 
we have found that strongly-disordered, weakly-coupled superconducting thin films are excellent superinductors with $L_{k/\square}$ enhanced up to two orders of magnitude by disorder while $Q \gtrsim 10^4$.
Moreover, close to the superconductor-insulator transition $Q$ has a broad probability distribution and, counterintuitively, is enhanced by disorder.
Despite the limitations mentioned previously, the enhancement of $L_{k/\square}$ with disorder and the mild decrease, and even enhancement in some cases, of $Q$ with disorder, are robust, can be confirmed experimentally and be used to improve the superinductor performance.
\phantomsection
\tocless 
\acknowledgments{ 
		B.F. and A.M.G.G. acknowledge financial support from a Shanghai talent program, from the National Natural Science Foundation China (NSFC) (Grant No. 11874259) and from the National Key R\&D Program of China (Project ID: 2019YFA0308603). A.S. acknowledges the computational facilities of Physics Department, Technion.}
	

\let\oldaddcontentsline\addcontentsline
\renewcommand{\addcontentsline}[3]{}
\bibliographystyle{unsrt}
\bibliography{libsinductor2}

\begin{thebibliography}{10}

\bibitem{you2002}
J.~Q. You, J.~S. Tsai, and Franco Nori.
\newblock Scalable quantum computing with josephson charge qubits.
\newblock {\em Phys. Rev. Lett.}, 89:197902, Oct 2002.

\bibitem{gu2017microwave}
Xiu Gu, Anton~Frisk Kockum, Adam Miranowicz, Yu-xi Liu, and Franco Nori.
\newblock Microwave photonics with superconducting quantum circuits.
\newblock {\em Physics Reports}, 718:1--102, 2017.

\bibitem{kjaergaard2020}
Morten Kjaergaard, Mollie~E Schwartz, Jochen Braum{\"u}ller, Philip Krantz,
  Joel I-J Wang, Simon Gustavsson, and William~D Oliver.
\newblock Superconducting qubits: Current state of play.
\newblock {\em Annual Review of Condensed Matter Physics}, 11:369--395, 2020.

\bibitem{peltonen2013}
J.~T. Peltonen, O.~V. Astafiev, Yu.~P. Korneeva, B.~M. Voronov, A.~A. Korneev,
  I.~M. Charaev, A.~V. Semenov, G.~N. Golt'sman, L.~B. Ioffe, T.~M. Klapwijk,
  and J.~S. Tsai.
\newblock Coherent flux tunneling through nbn nanowires.
\newblock {\em Phys. Rev. B}, 88:220506, Dec 2013.

\bibitem{manuchar2009}
Vladimir~E. Manucharyan, Jens Koch, Leonid~I. Glazman, and Michel~H. Devoret.
\newblock Fluxonium: Single cooper-pair circuit free of charge offsets.
\newblock {\em Science}, 326(5949):113--116, 2009.

\bibitem{annunziata2010}
Anthony~J Annunziata, Daniel~F Santavicca, Luigi Frunzio, Gianluigi Catelani,
  Michael~J Rooks, Aviad Frydman, and Daniel~E Prober.
\newblock Tunable superconducting nanoinductors.
\newblock {\em Nanotechnology}, 21(44):445202, oct 2010.

\bibitem{niepce2019}
David Niepce, Jonathan Burnett, and Jonas Bylander.
\newblock High kinetic inductance $\mathrm{Nb}\mathrm{N}$ nanowire
  superinductors.
\newblock {\em Phys. Rev. Applied}, 11:044014, Apr 2019.

\bibitem{zhang2019}
Wenyuan Zhang, K.~Kalashnikov, Wen-Sen Lu, P.~Kamenov, T.~DiNapoli, and M.E.
  Gershenson.
\newblock Microresonators fabricated from high-kinetic-inductance aluminum
  films.
\newblock {\em Phys. Rev. Applied}, 11:011003, Jan 2019.

\bibitem{rotzinger2016}
H~Rotzinger, S~T Skacel, M~Pfirrmann, J~N Voss, J~Münzberg, S~Probst,
  P~Bushev, M~P Weides, A~V Ustinov, and J~E Mooij.
\newblock Aluminium-oxide wires for superconducting high kinetic inductance
  circuits.
\newblock {\em Superconductor Science and Technology}, 30(2):025002, nov 2017.

\bibitem{valenti2019}
Francesco Valenti, Fabio Henriques, Gianluigi Catelani, Nataliya Maleeva, Lukas
  Grunhaupt, Uwe von Lupke, Sebastian~T. Skacel, Patrick Winkel, Alexander
  Bilmes, Alexey~V. Ustinov, Johannes Goupy, Martino Calvo, Alain Beno\^{\i}t,
  Florence Levy-Bertrand, Alessandro Monfardini, and Ioan~M. Pop.
\newblock Interplay between kinetic inductance, nonlinearity, and quasiparticle
  dynamics in granular aluminum microwave kinetic inductance detectors.
\newblock {\em Phys. Rev. Applied}, 11:054087, May 2019.

\bibitem{lukas2018}
Lukas Gr\"unhaupt, Nataliya Maleeva, Sebastian~T. Skacel, Martino Calvo,
  Florence Levy-Bertrand, Alexey~V. Ustinov, Hannes Rotzinger, Alessandro
  Monfardini, Gianluigi Catelani, and Ioan~M. Pop.
\newblock Loss mechanisms and quasiparticle dynamics in superconducting
  microwave resonators made of thin-film granular aluminum.
\newblock {\em Phys. Rev. Lett.}, 121:117001, Sep 2018.

\bibitem{maleeva2018}
N.~Maleeva, L.~Grünhaupt, T.~Klein, F.~Levy-Bertrand, O.~Dupre, M.~Calvo,
  F.~Valenti, P.~Winkel, F.~Friedrich, W.~Wernsdorfer, and et~al.
\newblock Circuit quantum electrodynamics of granular aluminum resonators.
\newblock {\em Nature Communications}, 9(1), Sep 2018.

\bibitem{lukas2019}
Lukas Grunhaupt, Martin Spiecker, Daria Gusenkova, Nataliya Maleeva,
  Sebastian~T. Skacel, Ivan Takmakov, Francesco Valenti, Patrick Winkel, Hannes
  Rotzinger, Wolfgang Wernsdorfer, and et~al.
\newblock Granular aluminium as a superconducting material for high-impedance
  quantum circuits.
\newblock {\em Nature Materials}, 18(8):816--819, Apr 2019.

\bibitem{deutscher2020}
Aviv Glezer~Moshe, Eli Farber, and Guy Deutscher.
\newblock Granular superconductors for high kinetic inductance and low loss
  quantum devices.
\newblock {\em Applied Physics Letters}, 117(6):062601, Aug 2020.

\bibitem{winkel2020}
Patrick Winkel, Kiril Borisov, Lukas Gr\"unhaupt, Dennis Rieger, Martin
  Spiecker, Francesco Valenti, Alexey~V. Ustinov, Wolfgang Wernsdorfer, and
  Ioan~M. Pop.
\newblock Implementation of a transmon qubit using superconducting granular
  aluminum.
\newblock {\em Phys. Rev. X}, 10:031032, Aug 2020.

\bibitem{fang2020}
Fang Yang, Thomas Gozlinski, Tim Storbeck, Lukas Gr\"unhaupt, Ioan~M. Pop, and
  Wulf Wulfhekel.
\newblock Microscopic charging and in-gap states in superconducting granular
  aluminum.
\newblock {\em Phys. Rev. B}, 102:104502, Sep 2020.

\bibitem{dupre2017}
O~Dupré, A~Benoît, M~Calvo, A~Catalano, J~Goupy, C~Hoarau, T~Klein, K~Le
  Calvez, B~Sacépé, A~Monfardini, and et~al.
\newblock Tunable sub-gap radiation detection with superconducting resonators.
\newblock {\em Superconductor Science and Technology}, 30(4):045007, Feb 2017.

\bibitem{de2018}
SE~De~Graaf, ST~Skacel, T~H{\"o}nigl-Decrinis, R~Shaikhaidarov, H~Rotzinger,
  S~Linzen, M~Ziegler, U~H{\"u}bner, H-G Meyer, V~Antonov, et~al.
\newblock Charge quantum interference device.
\newblock {\em Nature Physics}, 14(6):590--594, 2018.

\bibitem{coumou2013}
P.~C. J.~J. Coumou, E.~F.~C. Driessen, J.~Bueno, C.~Chapelier, and T.~M.
  Klapwijk.
\newblock Electrodynamic response and local tunneling spectroscopy of strongly
  disordered superconducting tin films.
\newblock {\em Phys. Rev. B}, 88:180505, Nov 2013.

\bibitem{Deutscher1973}
G.~Deutscher, H.~Fenichel, M.~Gershenson, E.~Gr\"{u}nbaum, and Z.~Ovadyahu.
\newblock {Transition to zero dimensionality in granular aluminum
  superconducting films}.
\newblock {\em J. Low Temp. Phys.}, 10(1-2):231--243, January 1973.

\bibitem{Deutscher1973a}
G.~Deutscher, M.~Gershenson, E.~Gr\"{u}nbaum, and Y.~Imry.
\newblock {Granular Superconducting Films}.
\newblock {\em J. Vac. Sci. Technol.}, 10(5):697, September 1973.

\bibitem{Abeles1977}
B~Abeles.
\newblock {Effect of charging energy on superconductivity in granular metal
  films}.
\newblock {\em Phys. Rev. B}, 15(5):2828, 1977.

\bibitem{Abeles1966}
B.~Abeles, Roger~W. Cohen, and G.~W. Cullen.
\newblock {Enhancement of Superconductivity in Metal Films}.
\newblock {\em Phys. Rev. Lett.}, 17(12):632--634, September 1966.

\bibitem{meservey1969}
Robert Meservey and Paul~M Tedrow.
\newblock Measurements of the kinetic inductance of superconducting linear
  structures.
\newblock {\em Journal of Applied Physics}, 40(5):2028--2034, 1969.

\bibitem{anderson1958sc}
P.~W. Anderson.
\newblock Random-phase approximation in the theory of superconductivity.
\newblock {\em Phys. Rev.}, 112:1900--1916, Dec 1958.

\bibitem{anderson1963}
P.~W. Anderson.
\newblock Plasmons, gauge invariance, and mass.
\newblock {\em Phys. Rev.}, 130:439--442, Apr 1963.

\bibitem{mattis1958}
D.~C. Mattis and J.~Bardeen.
\newblock Theory of the anomalous skin effect in normal and superconducting
  metals.
\newblock {\em Phys. Rev.}, 111:412--417, Jul 1958.

\bibitem{goldstone1961}
Jeffrey Goldstone.
\newblock Field theories with {\guillemotleft}superconductor{\guillemotright}
  solutions.
\newblock {\em Il Nuovo Cimento (1955-1965)}, 19(1):154--164, 1961.

\bibitem{seibold2017}
G.~Seibold, L.~Benfatto, and C.~Castellani.
\newblock Application of the mattis-bardeen theory in strongly disordered
  superconductors.
\newblock {\em Phys. Rev. B}, 96:144507, Oct 2017.

\bibitem{cea2014}
T.~Cea, D.~Bucheli, G.~Seibold, L.~Benfatto, J.~Lorenzana, and C.~Castellani.
\newblock Optical excitation of phase modes in strongly disordered
  superconductors.
\newblock {\em Phys. Rev. B}, 89:174506, May 2014.

\bibitem{cea2015}
T.~Cea, C.~Castellani, G.~Seibold, and L.~Benfatto.
\newblock Nonrelativistic dynamics of the amplitude (higgs) mode in
  superconductors.
\newblock {\em Phys. Rev. Lett.}, 115:157002, Oct 2015.

\bibitem{abhisek2020}
Abhisek Samanta, Amulya Ratnakar, Nandini Trivedi, and Rajdeep Sensarma.
\newblock Two-particle spectral function for disordered $\mathit{s}$-wave
  superconductors: Local maps and collective modes.
\newblock {\em Phys. Rev. B}, 101:024507, Jan 2020.

\bibitem{bofan2021a}
Bo~Fan, Abhisek Samanta, and Antonio~M. Garc\'{\i}a-Garc\'{\i}a.
\newblock Characterization of collective excitations in weakly coupled
  disordered superconductors.
\newblock {\em Phys. Rev. B}, 105:094515, Mar 2022.

\bibitem{martinis2005}
John~M. Martinis, K.~B. Cooper, R.~McDermott, Matthias Steffen, Markus Ansmann,
  K.~D. Osborn, K.~Cicak, Seongshik Oh, D.~P. Pappas, R.~W. Simmonds, and
  Clare~C. Yu.
\newblock Decoherence in josephson qubits from dielectric loss.
\newblock {\em Phys. Rev. Lett.}, 95:210503, Nov 2005.

\bibitem{gao2008}
Jiansong Gao.
\newblock The physics of superconducting microwave resonators. dissertation
  (ph.d.), california institute of technology, 2008.

\bibitem{Mayoh2015}
James Mayoh and Antonio~M. Garc\'{\i}a-Garc\'{\i}a.
\newblock Global critical temperature in disordered superconductors with weak
  multifractality.
\newblock {\em Phys. Rev. B}, 92:174526, Nov 2015.

\bibitem{bofan2020}
Bo~Fan and Antonio~M. Garc\'{\i}a-Garc\'{\i}a.
\newblock Enhanced phase-coherent multifractal two-dimensional
  superconductivity.
\newblock {\em Phys. Rev. B}, 101:104509, Mar 2020.

\bibitem{bofan2020a}
Bo~Fan and Antonio~M. Garc\'{\i}a-Garc\'{\i}a.
\newblock Superconductivity at the three-dimensional anderson metal-insulator
  transition.
\newblock {\em Phys. Rev. B}, 102:184507, Nov 2020.

\bibitem{Burmistrov2012}
I.~S. Burmistrov, I.~V. Gornyi, and A.~D. Mirlin.
\newblock {Enhancement of the Critical Temperature of Superconductors by
  Anderson Localization}.
\newblock {\em Phys. Rev. Lett.}, 108(1):017002, January 2012.

\bibitem{Mayoh2014}
James Mayoh and Antonio~M. Garc\'{\i}a-Garc\'{\i}a.
\newblock {Number theory, periodic orbits, and superconductivity in nanocubes}.
\newblock {\em Phys. Rev. B}, 90(1):014509, July 2014.

\bibitem{Mayoh2014a}
J.~Mayoh and A.~M. Garc\'{\i}a-Garc\'{\i}a.
\newblock {Strong enhancement of bulk superconductivity by engineered
  nanogranularity}.
\newblock {\em Phys. Rev. B}, 90(13):134513, October 2014.

\bibitem{Burmistrov2015}
I.~S. Burmistrov, I.~V. Gornyi, and A.~D. Mirlin.
\newblock Superconductor-insulator transitions: Phase diagram and
  magnetoresistance.
\newblock {\em Phys. Rev. B}, 92:014506, Jul 2015.

\bibitem{tezuka2010}
Masaki Tezuka and Antonio~M. García-García.
\newblock Stability of the superfluid state in a disordered one-dimensional
  ultracold fermionic gas.
\newblock {\em Physical Review A}, 82(4), Oct 2010.

\bibitem{Feigelman2007}
M.~V. Feigel'man, L.~B. Ioffe, V.~E. Kravtsov, and E.~A. Yuzbashyan.
\newblock Eigenfunction fractality and pseudogap state near the
  superconductor-insulator transition.
\newblock {\em Phys. Rev. Lett.}, 98:027001, Jan 2007.

\bibitem{Parmenter1968a}
R.H. Parmenter.
\newblock {Size Effect in a Granular Superconductor}.
\newblock {\em Phys. Rev.}, 166(2):392--396, February 1968.

\bibitem{Shanenko2007}
A.~A. Shanenko, M.~D. Croitoru, and F.~M. Peeters.
\newblock {Oscillations of the superconducting temperature induced by quantum
  well states in thin metallic films: Numerical solution of the Bogoliubov-de
  Gennes equations}.
\newblock {\em Phys. Rev. B}, 75(1):014519, January 2007.

\bibitem{garcia2008bardeen}
Antonio~M Garc{\'\i}a-Garc{\'\i}a, Juan~Diego Urbina, Emil~A Yuzbashyan, Klaus
  Richter, and Boris~L Altshuler.
\newblock Bardeen-cooper-schrieffer theory of finite-size superconducting
  metallic grains.
\newblock {\em Physical review letters}, 100(18):187001, 2008.

\bibitem{Garcia-Garcia2011}
Antonio~M. Garc\'{\i}a-Garc\'{\i}a, Juan~D Urbina, Emil~A. Yuzbashyan, Klaus
  Richter, and Boris~L. Altshuler.
\newblock {BCS superconductivity in metallic nanograins: Finite-size
  corrections, low-energy excitations, and robustness of shell effects}.
\newblock {\em Phys. Rev. B}, 83(1):014510, January 2011.

\bibitem{brun2014}
C.~Brun, T.~Cren, V.~Cherkez, F.~Debontridder, S.~Pons, D.~Fokin, M.~C.
  Tringides, S.~Bozhko, L.~B. Ioffe, B.~L. Altshuler, et~al.
\newblock Remarkable effects of disorder on superconductivity of single atomic
  layers of lead on silicon.
\newblock {\em Nature Physics}, 10:444--450, April 2014.

\bibitem{xue2019}
Kun Zhao, Haicheng Lin, Xiao Xiao, Wantong Huang, Wei Yao, Mingzhe Yan, Ying
  Xing, Qinghua Zhang, Zi-Xiang Li, Shintaro Hoshino, et~al.
\newblock Disorder induced multifractal superconductivity in monolayer niobium
  dichalcogenides.
\newblock {\em arXiv preprint arXiv:1904.07076 Nat. Phys. 15, 904}, 2019.

\bibitem{verdu2018}
Carmen Rubio-Verdu, Antonio~M. Garcia-Garcia, Hyejin Ryu, Deung-Jang Choi,
  Javier Zaldivar, Shujie Tang, Bo~Fan, Zhi-Xun Shen, Sung-Kwan Mo,
  Jose~Ignacio Pascual, and Miguel~M. Ugeda.
\newblock Visualization of multifractal superconductivity in a two-dimensional
  transition metal dichalcogenide in the weak-disorder regime.
\newblock {\em Nano Letters}, 20(7):5111--5118, 2020.

\bibitem{pracht2016}
Uwe~S. Pracht, Nimrod Bachar, Lara Benfatto, Guy Deutscher, Eli Farber, Martin
  Dressel, and Marc Scheffler.
\newblock Enhanced cooper pairing versus suppressed phase coherence shaping the
  superconducting dome in coupled aluminum nanograins.
\newblock {\em Phys. Rev. B}, 93:100503, Mar 2016.

\bibitem{pracht2017}
Uwe~S. Pracht, Tommaso Cea, Nimrod Bachar, Guy Deutscher, Eli Farber, Martin
  Dressel, Marc Scheffler, Claudio Castellani, Antonio~M.
  Garc\'{\i}a-Garc\'{\i}a, and Lara Benfatto.
\newblock Optical signatures of the superconducting goldstone mode in granular
  aluminum: Experiments and theory.
\newblock {\em Phys. Rev. B}, 96:094514, Sep 2017.

\bibitem{Bose2010}
Sangita Bose, Antonio~M. Garc\'{\i}a-Garc\'{\i}a, Miguel~M. Ugeda, Juan~D.
  Urbina, Christian~H. Michaelis, Ivan Brihuega, and Klaus Kern.
\newblock {Observation of shell effects in superconducting nanoparticles of
  Sn.}
\newblock {\em Nat. Mater.}, 9(7):550--554, July 2010.

\bibitem{li2015}
Zhi Li, Jun-Ping Peng, Hui-Min Zhang, Can-Li Song, Shuai-Hua Ji, Lili Wang,
  Ke~He, Xi~Chen, Qi-Kun Xue, and Xu-Cun Ma.
\newblock Visualizing superconductivity in fese nanoflakes on
  ${\mathrm{srtio}}_{3}$ by scanning tunneling microscopy.
\newblock {\em Phys. Rev. B}, 91:060509, Feb 2015.

\bibitem{Brihuega2011}
Ivan Brihuega, Pedro Ribeiro, Antonio~M. Garcia-Garcia, Miguel~M. Ugeda,
  Christian~H. Michaelis, Sangita Bose, Klaus Kern, and Antonio
  Garc\'{\i}a-Garc\'{\i}a.
\newblock {Experimental observation of thermal fluctuations in single
  superconducting Pb nanoparticles through tunneling measurements}.
\newblock {\em Phys. Rev. B}, 84(10):104525, April 2011.

\bibitem{DeGennes1964}
P.G. de~Gennes.
\newblock {Boundary Effects in Superconductors}.
\newblock {\em Rev. Mod. Phys.}, 36(1):225--237, January 1964.

\bibitem{DeGennes1966}
P.G. de~Gennes.
\newblock {\em {Superconductivity of Metals and Alloys}}.
\newblock W.A. Bebjamin, inc., New York, 1966.

\bibitem{ioffe2018}
M.~V. Feigel'man and L.~B. Ioffe.
\newblock Microwave properties of superconductors close to the
  superconductor-insulator transition.
\newblock {\em Phys. Rev. Lett.}, 120:037004, Jan 2018.

\bibitem{hafner2014surface}
Daniel Hafner, Martin Dressel, and Marc Scheffler.
\newblock Surface-resistance measurements using superconducting stripline
  resonators.
\newblock {\em Review of Scientific Instruments}, 85(1):014702, 2014.

\bibitem{suppl}
See supplemental material at \url{http://link.aps.org/
  supplemental/10.1103/PhysRevLett.130.047001} for discussions about: (i) the
  theoretical model and the technical details related to the calculation of
  quality factor and kinetic inductance in disordered superconductors, (ii)
  temperature dependence of the sample-averaged quality factor and kinetic
  inductance for different disorder strengths, (iii) the quality factor for a
  cavity, and (iv) the estimation of critical disorder using the spectral gap
  and a percolation analysis.

\bibitem{yassin1995electromagnetic}
Ghassan Yassin and S~Withington.
\newblock Electromagnetic models for superconducting millimetre-wave and
  sub-millimetre-wave microstrip transmission lines.
\newblock {\em Journal of Physics D: Applied Physics}, 28(9):1983, 1995.

\bibitem{devisser2014}
P.~J. de~Visser, D.~J. Goldie, P.~Diener, S.~Withington, J.~J.~A. Baselmans,
  and T.~M. Klapwijk.
\newblock Evidence of a nonequilibrium distribution of quasiparticles in the
  microwave response of a superconducting aluminum resonator.
\newblock {\em Phys. Rev. Lett.}, 112:047004, Jan 2014.

\bibitem{cheng1989field}
David~Keun Cheng et~al.
\newblock {\em Field and wave electromagnetics}.
\newblock Pearson Education India, 1989.

\bibitem{Ghosal2001}
Amit Ghosal, Mohit Randeria, and Nandini Trivedi.
\newblock {Inhomogeneous pairing in highly disordered s-wave superconductors}.
\newblock {\em Phys. Rev. B}, 65(1):014501, November 2001.

\bibitem{ghosal1998}
Amit Ghosal, Mohit Randeria, and Nandini Trivedi.
\newblock Role of spatial amplitude fluctuations in highly disordered
  $\mathit{s}$-wave superconductors.
\newblock {\em Phys. Rev. Lett.}, 81:3940--3943, Nov 1998.

\bibitem{datta2021}
Anushree Datta, Anurag Banerjee, Nandini Trivedi, and Amit Ghosal.
\newblock New paradigm for a disordered superconductor in a magnetic field.
\newblock 2021.

\bibitem{huscroft1998}
Carey Huscroft and Richard~T. Scalettar.
\newblock Evolution of the density of states gap in a disordered
  superconductor.
\newblock {\em Phys. Rev. Lett.}, 81:2775--2778, Sep 1998.

\bibitem{grunhaupt2019granular}
Lukas Gr{\"u}nhaupt, Martin Spiecker, Daria Gusenkova, Nataliya Maleeva,
  Sebastian~T Skacel, Ivan Takmakov, Francesco Valenti, Patrick Winkel, Hannes
  Rotzinger, Wolfgang Wernsdorfer, et~al.
\newblock Granular aluminium as a superconducting material for high-impedance
  quantum circuits.
\newblock {\em Nature materials}, 18(8):816--819, 2019.

\bibitem{krantz2019quantum}
Philip Krantz, Morten Kjaergaard, Fei Yan, Terry~P Orlando, Simon Gustavsson,
  and William~D Oliver.
\newblock A quantum engineer's guide to superconducting qubits.
\newblock {\em Applied Physics Reviews}, 6(2):021318, 2019.

\bibitem{sergeev2002ultrasensitive}
AV~Sergeev, VV~Mitin, and BS~Karasik.
\newblock Ultrasensitive hot-electron kinetic-inductance detectors operating
  well below the superconducting transition.
\newblock {\em Applied physics letters}, 80(5):817--819, 2002.

\bibitem{samkharadze2016high}
Nodar Samkharadze, A~Bruno, Pasquale Scarlino, G~Zheng, DP~DiVincenzo,
  L~DiCarlo, and LMK Vandersypen.
\newblock High-kinetic-inductance superconducting nanowire resonators for
  circuit qed in a magnetic field.
\newblock {\em Physical Review Applied}, 5(4):044004, 2016.

\bibitem{chang1977}
Jhy-Jiun Chang and D.~J. Scalapino.
\newblock Kinetic-equation approach to nonequilibrium superconductivity.
\newblock {\em Phys. Rev. B}, 15:2651--2670, Mar 1977.

\bibitem{wolter1981}
R.E. Horstman and J.~Wolter.
\newblock Gap enhancement in narrow superconducting tunneljunctions induced by
  homogeneous microwave currents.
\newblock {\em Physics Letters A}, 82(1):43--45, 1981.

\bibitem{tinkham2004}
Michael Tinkham.
\newblock {\em Introduction to superconductivity}.
\newblock Courier Corporation, 2004.

\bibitem{peruzzo2020}
M.~Peruzzo, A.~Trioni, F.~Hassani, M.~Zemlicka, and J.~M. Fink.
\newblock Surpassing the resistance quantum with a geometric superinductor.
\newblock {\em Phys. Rev. Applied}, 14:044055, Oct 2020.

\bibitem{mayoh2015global}
James Mayoh and Antonio~M. Garc\'{\i}a-Garc\'{\i}a.
\newblock Global critical temperature in disordered superconductors with weak
  multifractality.
\newblock {\em Phys. Rev. B}, 92:174526, Nov 2015.

\bibitem{nambu1960}
Yoichiro Nambu.
\newblock Quasi-particles and gauge invariance in the theory of
  superconductivity.
\newblock {\em Phys. Rev.}, 117:648--663, Feb 1960.

\bibitem{waldram1976}
JR~Waldram.
\newblock The josephson effects in weakly coupled superconductors.
\newblock {\em Reports on Progress in Physics}, 39(8):751, 1976.

\bibitem{glezer2021}
Aviv Glezer~Moshe, Gal Tuvia, Shilo Avraham, Eli Farber, and Guy Deutscher.
\newblock Tunneling study in granular aluminum near the mott metal-to-insulator
  transition.
\newblock {\em Physical Review B}, 104(5), Aug 2021.

\bibitem{seibold2015}
G.~Seibold, L.~Benfatto, C.~Castellani, and J.~Lorenzana.
\newblock Amplitude, density, and current correlations of strongly disordered
  superconductors.
\newblock {\em Phys. Rev. B}, 92:064512, Aug 2015.

\bibitem{kamenov2020granular}
Plamen Kamenov, Wen-Sen Lu, Konstantin Kalashnikov, Thomas DiNapoli, Matthew~T
  Bell, and Michael~E Gershenson.
\newblock Granular aluminum meandered superinductors for quantum circuits.
\newblock {\em Physical Review Applied}, 13(5):054051, 2020.

\bibitem{stauffer2003introduction}
D~Stauffer and A~Aharony.
\newblock Introduction to percolation theory (2003).
\newblock {\em London: Taylor \& Francis}, 2003.

\bibitem{dean1967monte}
P~Dean and NF~Bird.
\newblock Monte carlo estimates of critical percolation probabilities.
\newblock In {\em Mathematical Proceedings of the Cambridge Philosophical
  Society}, volume~63, pages 477--479. Cambridge University Press, 1967.

\end{thebibliography}
\let\addcontentsline\oldaddcontentsline
\onecolumngrid
\clearpage

	\allowdisplaybreaks[2]

	\renewcommand{\thefigure}{S\arabic{figure}}
	\setcounter{figure}{0}
	\renewcommand{\theequation}{S\arabic{equation}}
	\setcounter{equation}{0}
	\renewcommand\thesection{S\arabic{section}}
	\setcounter{section}{0}

	\title{Supplemental Material for ``Tuning superinductors by quantum coherence effects for enhancing quantum computing"}

	\begin{abstract}
		In this supplemental information, we first introduce the theoretical model and present the technical and numerical details related to the calculation of the quality factor and the kinetic inductance at finite temperatures. Secondly, we show that around the critical disorder, finite size effects are not important in our analysis. 
		Next, we present the temperature-dependent quality factor for a series of disorder strengths and frequencies by first computing the averaged conductivity. We also show that these results are similar to those obtained by computing the quality factor for each disorder realization and then performing ensemble average. The quality factor in the cavity case is also discussed for comparison.
		Finally, we make an estimation of the critical disorder at which the superconductor transition occurs by a percolation analysis and also by a study of the dependence of the spectroscopic gap with the disorder strength. 
	\end{abstract}
	\maketitle
	\onecolumngrid
	\tableofcontents
	
	\section{Calculation of the quality factor and the kinetic inductance from a gauge-invariant optical conductivity at finite temperature}
	The microscopic Hamiltonian is an attractive Hubbard model on a square lattice in the presence of onsite disorder,
	\begin{equation}
		H = -t\sum_{\langle ij\rangle\sigma} c^\dagger_{i\sigma} c_{j\sigma} - U\sum_i n_{i\uparrow}n_{i\downarrow} + \sum_i V_i n_i
	\end{equation} 
	where $t$ is the nearest neighbor hopping energy and $U$ is the attractive interaction between two electrons on the same site which leads to the Cooper pairing between them. The onsite random potential is drawn from a uniform distribution of zero mean and width $2V$, i.e. $V_i\in [-V,V]$ where $V$ is the strength of the disordered potential. Throughout the paper, we have used the values of $U$ and $V$ in units of $t$.
	
	The goal of this section is to express the quality factor ($Q$) and kinetic inductance ($L_{k/\square}$), introduced in the main paper, in terms of the finite temperature conductivity, which includes vertex corrections to restore gauge invariance. Our starting point is the mean-field limit of this Hamiltonian, the so called BdG (Bogoliubov de-Gennes) Hamiltonian~\cite{DeGennes1964,DeGennes1966}, which depends on two parameters, the local superconducting order parameter $\Delta(i)$ and the local density $n(i)$~\cite{ghosal1998,Ghosal2001}. We then invoke a canonical transformation (Bogoliubov transformation) to diagonalize the effective mean-field Hamiltonian. Note that since we are working in presence of random potential, the translational invariance of the system is broken and we need to diagonalize the large BdG Hamiltonian in real space. We obtain the eigenvalues $E_n$ and eigenfunctions $\{u_n(i),v_n(i)\}$. With this information, we solve the self-consistent BdG mean-field equations iteratively on each site of the square lattice.
	
	The next step is to use these mean-field results to study the optical response of the system at finite temperatures by an explicit calculation of the conductivity. This calculation includes corrections to the mean field results which are expressed in terms of the solutions of the BdG equation.\\
	The conductivity depends on the following dynamical correlation function \cite{cea2014,cea2015,seibold2015,seibold2017},
	\begin{equation}
		\chi_{ij}(\phi,\phi')=-i \int dt e^{i\omega t} \langle \left[\phi_i(t),\ \phi'_j(0)\right] \rangle
		\label{eq.cor_fun}
	\end{equation} 
	where $\phi$ stands for the fluctuation components and the current operators given by \cite{cea2014}
	\begin{equation}
		\begin{aligned}
			&\delta\Delta_i &=\ &c_{i\downarrow}c_{i\uparrow} - \langle c_{i\downarrow}c_{i\uparrow} \rangle \\
			&\delta\Delta_i^\dagger &=\ &c_{i\uparrow}^\dagger c_{i\downarrow}^\dagger - \langle c_{i\uparrow}^\dagger c_{i\downarrow}^\dagger \rangle \\
			&\delta n_i &=\ &\sum_{\sigma} \left( c_{i\sigma}^\dagger c_{i\sigma} - \langle c_{i\sigma}^\dagger c_{i\sigma} \rangle \right) \\
			&j_i^\alpha &=\ &it \sum_{\sigma} \left( c_{i+\alpha,\sigma}^\dagger c_{i\sigma} - c_{i\sigma}^\dagger c_{i+\alpha,\sigma} \right). \\
		\end{aligned}
		\label{eq.fluct_curr}
	\end{equation} 
	Here $\delta\Delta_i$ is the fluctuation in the local superconducting order parameter around its mean-field value of $\Delta(i)$, and $\delta n_i$ is the fluctuation in local density around its mean-field value of $n(i)$. $\langle\cdots\rangle$ denotes the expectation value of the operator in the inhomogeneous BdG eigenstate (obtained by diagonalizing the BdG matrix). The amplitude fluctuation $A_i$ and the phase fluctuation $\Phi_i$ of the superconducting order parameter $\Delta(i)$ are therefore given by 
	\begin{align}
		A_i=(\delta\Delta_i+\delta\Delta_i^\dagger)/\sqrt{2} \label{eq.amp}\\
		\Phi_i=i(\delta\Delta_i-\delta\Delta_i^\dagger)/\sqrt{2}
		\label{eq.pha}
	\end{align}
	Note that in our case, all the eigenvalues $E_n$ and the corresponding eigenvectors $\{u_n(i),v_n(i)\}$ of the BdG matrix are real. However here we present the formulas for different correlation functions at finite temperature $T$ for the general case, where the eigenvectors are complex numbers.
	
	The bare current-current correlation function \cite{cea2014,cea2015,seibold2015,seibold2017} is given by,  
	\begin{equation}
		\begin{aligned}
			\chi_{ij}(j^x,j^x) = -2t^2 \sum_{nm} &
			{u_n^*(i+\hat{x})u_m(i) (u_{m}^*(j+\hat{x})u_{n}(j) - v_n(j+\hat{x})v_m^*(j) ) \over \omega + i\eta + E_n-E_m } (f(E_n) - f(E_m))\\
			&+ {u_n^*(i+\hat{x})v_m^*(i) ( v_{m}(j+\hat{x})u_{n}(j) + v_{n}(j+\hat{x})u_{m}(j) ) \over \omega + i\eta + E_n+E_m } (f(E_n) + f(E_m) - 1)\\
			&+ { v_n(i+\hat{x})u_m(i) ( u_{m}^*(j+\hat{x})v_{n}^*(j) + u_{n}^*(j+\hat{x})v_{m}^*(j) ) \over \omega + i\eta -E_n-E_m } (1 - f(E_n) - f(E_m))\\
			&+ { v_n(i+\hat{x})v_m^*(i) ( u_{n}^*(j+\hat{x})u_{m}(j) - v_{n}(j+\hat{x})v_{m}^*(j) ) \over \omega + i\eta -E_n+E_m }(-f(E_n)+f(E_m))\\
			&-(j+\hat{x}\leftrightarrow j) -(i+\hat{x}\leftrightarrow i) +(i+\hat{x}\leftrightarrow i,j+\hat{x}\leftrightarrow j)
		\end{aligned}
		\label{eq.A8}
	\end{equation} 
	where $f(E_n) = \frac{1}{e^{E_n/T}+1}$ is the Fermi-Dirac function. Similarly, the correlation function of the current operator $j^x$ and the pair fluctuations ($\delta\Delta$, $\delta\Delta^\dagger$) or the charge fluctuations ($\delta n$) are given by,
	\begin{align}
		\nonumber \chi_{ij}(j^x,\delta \Delta) =~ &2it\sum_{nm}  \frac{\left[u_n^*(i+\hat{x})u_m(i) - u_n^*(i)u_m(i+\hat{x})\right] u_{n}(j)v_{m}^*(j)}{\omega + i\eta + E_n-E_m}\ (f(E_n)-f(E_m))\ \\
		\nonumber & -\frac{\left[u_n^*(i+\hat{x})v_m^*(i) - u_n^*(i)v_m^*(i+\hat{x})\right] u_{m}(j)u_{n}(j)}{\omega + i\eta + E_n+E_m}\ (f(E_n)+f(E_m)-1) \\
		\nonumber & +\frac{\left[v_n(i+\hat{x})u_m(i) - v_n(i)u_m(i+\hat{x})\right] v_{m}^*(j) v_{n}^*(j)}{\omega + i\eta - E_n-E_m}\ (1-f(E_n)-f(E_m)) \\
		& -\frac{\left[v_n(i+\hat{x})v_m^*(i) - v_n(i)v_m^*(i+\hat{x})\right] v_{n}^*(j)u_{m}(j)}{\omega + i\eta - E_n+E_m}\ (-f(E_n)+f(E_m)) \\
		\no\\
		\nonumber \chi_{ij}(j^x,\delta \Delta^\dagger) =~ & 2it\sum_{nm} \frac{ [u_n^*(i+\hat{x})u_m(i) - u_n^*(i)u_m(i+\hat{x})] u_{m}^*(j)v_{n}(j)}{\omega+i\eta +E_n-E_m} (f(E_n)-f(E_m)) \\
		\nonumber &+\frac{ [ u_n^*(i+\hat{x})v_m^*(i) - u_n^*(i)v_m^*(i+\hat{x}) ]  v_{m}(j)v_{n}(j)}{\omega+i\eta +E_n+E_m} (f(E_n)+f(E_m)-1) \\
		\nonumber &-\frac{ [ v_n(i+\hat{x})u_m(i) - v_n(i)u_m(i+\hat{x}) ] u_{m}^*(j)u_{n}^*(j) }{\omega+i\eta -E_n-E_m}(1-f(E_n)-f(E_m)) \\
		&-\frac{ [ v_n(i+\hat{x})v_m^*(i) - v_n(i)v_m^*(i+\hat{x}) ] u_{n}^*(j)v_{m}(j)}{\omega+i\eta -E_n+E_m} (-f(E_n)+f(E_m)) \\
		\no\\
		\nonumber \chi_{ij}(j^x,\delta n)=~& 2it\sum_{nm}  \frac{[u_n^*(i+\hat{x})u_m(i) - u_n^*(i)u_m(i+\hat{x})] [ u_{m}^*(j)u_{n}(j) - v_{n}(j)v_{m}^*(j) ]}{\omega + i\eta + E_n - E_m} (f(E_n)-f(E_m)) \\
		\nonumber &+ \frac{[u_n^*(i+\hat{x})v_m^*(i) - u_n^*(i)v_m^*(i+\hat{x})] [ v_{m}(j)u_{n}(j) + v_{n}(j)u_{m}(j) ]}{\omega + i\eta + E_n + E_m} (f(E_n)+f(E_m)-1) \\
		\nonumber &+ \frac{[ v_n(i+\hat{x})u_m(i) - v_n(i)u_m(i+\hat{x}) ] [ u_{m}^*(j)v_{n}^*(j) + u_{n}^*(j)v_{m}^*(j) ]}{\omega + i\eta - E_n - E_m}  (1-f(E_n)-f(E_m)) \\
		&+ \frac{[ v_n(i+\hat{x})v_m^*(i) - v_n(i)v_m^*(i+\hat{x}) ] [ v_{m}(j)v_{n}^*(j) - u_{n}^*(j)u_{m}(j) ]}{\omega + i\eta - E_n + E_m} (-f(E_n)+f(E_m)) 
	\end{align}
	The correlation functions between the pair fluctuations ($\delta \Delta$, $\delta \Delta^\dagger$) and the density fluctuations ($\delta n$) are given by, 
	\begin{align}
		\chi_{ij}(\delta \Delta,\delta \Delta) &= \sum_{nm}
		-\frac{u_n(i)u_m(i) v_{m}^*(j) v_{n}^*(j) }{ \omega + i\eta - E_n - E_m } ( 1 - f(E_n) - f(E_m) )
		+\frac{u_n(i)v_m^*(i) u_{m}(j)v_{n}^*(j) }{ \omega + i\eta - E_n + E_m }  (-f(E_n) + f(E_m)) \no\\
		&~~~~~~~+\frac{v_n^*(i)u_m(i)v_{m}^*(j)u_{n}(j) }{ \omega + i\eta + E_n - E_m } (f(E_n) - f(E_m))
		-\frac{v_n^*(i) v_m^*(i) u_{m}(j)u_{n}(j) }{ \omega + i\eta + E_n + E_m } ( f(E_n) + f(E_m) - 1 ) \\ 
		\no\\ 
		\chi_{ij}(\delta \Delta,\delta \Delta^\dagger) &= \sum_{nm}
		\frac{u_n(i)u_m(i)u_{m}^*(j)u_{n}^*(j)}{ \omega + i\eta -E_n - E_m } (1-f(E_m)-f(E_n)) 
		+ \frac{u_n(i)v_m^*(i)v_{m}(j)u_{n}^*(j)}{ \omega + i\eta - E_n + E_m } (-f(E_n) + f(E_m)) \no\\
		&~~~~~+ \frac{v_n^*(i)u_m(i)u_{m}^*(j)v_{n}(j)}{ \omega + i\eta + E_n - E_m } (f(E_n) - f(E_m))
		+ \frac{v_n^*(i) v_m^*(i) v_{m}(j)v_{n}(j)}{ \omega + i\eta + E_n + E_m} (f(E_n)+f(E_m)-1) \\
		\no\\
		\chi_{ij}(\delta \Delta^\dagger,\delta \Delta) &= \sum_{nm}
		\frac{u_n^*(i)u_m^*(i) u_{m}(j)u_{n}(j) }{ \omega + i\eta + E_n + E_m } (f(E_n)+f(E_n)-1)
		+ \frac{u_n^*(i)v_m(i) v_{m}^*(j)u_{n}(j) }{ \omega + i\eta + E_n - E_m } (f(E_n) - f(E_m)) \no\\
		&~~~~~+ \frac{v_n(i)u_m^*(i) u_{m}(j)v_{n}^*(j) }{ \omega + i\eta - E_n + E_m } (-f(E_n)+f(E_m)) 
		+ \frac{v_n(i)v_m(i) v_{m}^*(j) v_{n}^*(j) }{ \omega + i\eta - E_n - E_m } (1-f(E_m)-f(E_n)) 
	\end{align}
	\begin{align}
		\nonumber \chi_{ij}(\delta n,\delta \Delta) &= 2\sum_{nm} \frac{u_n^*(i)u_m(i) u_{n}(j)v_{m}^*(j)}{\omega+i\eta+E_n-E_m}\ (f(E_n)-f(E_m)) 
		-\frac{u_n^*(i)v_m^*(i) u_{m}(j)u_{n}(j)}{\omega+i\eta+E_n+E_m}\ (f(E_n)+f(E_m)-1) \no\\
		& ~~~~~+\frac{v_n(i)u_m(i) v_{m}^*(j) v_{n}^*(j)}{\omega+i\eta-E_n-E_m}\ (1-f(E_n)-f(E_m)) 
		-\frac{v_n(i)v_m^*(i) v_{n}^*(j)u_{m}(j)}{\omega+i\eta-E_n+E_m}\ (-f(E_n)+f(E_m))\ \\
		\no\\
		\chi_{ij}(\delta n,\delta \Delta^\dagger) &= 2\sum_{nm} \frac{u_n^*(i)u_m(i) u_{m}^*(j)v_{n}(j)}{\omega+i\eta+E_n-E_m} (f(E_n)-f(E_m))
		+\frac{u_n^*(i)v_m^*(i) v_{m}(j)v_{n}(j)}{\omega+i\eta+E_n+E_m} (f(E_n)+f(E_m)-1) \no\\
		& ~~~~~~~-\frac{v_n(i)u_m(i) u_{n}^*(j)u_{m}^*(j)}{\omega+i\eta-E_n-E_m} (1-f(E_n)-f(E_m))
		-\frac{v_n(i)v_m^*(i) u_{n}^*(j)v_{m}(j)}{\omega+i\eta-E_n+E_m} (-f(E_n)+f(E_m))
	\end{align}
	\begin{align}
		&\chi_{ij}(\delta n,\delta n) = \no\\
		&2\sum_{nm} \frac{u_n^*(i)u_m(i)[ u_{m}^*(j)u_{n}(j) - v_{n}(j)v_{m}^*(j) ]}{\omega+i\eta+E_n-E_m} (f(E_n)-f(E_m)) 
		+ \frac{u_n^*(i)v_m^*(i)[ v_{m}(j)u_{n}(j) + v_{n}(j)u_{m}(j) ]}{\omega+i\eta+E_n+E_m} (f(E_n)+f(E_m)-1) \no\\
		& + \frac{v_n(i)u_m(i)[ u_{m}^*(j)v_{n}^*(j) + u_{n}^*(j)v_{m}^*(j) ]}{\omega+i\eta-E_n-E_m} (1-f(E_n)-f(E_m))
		+ \frac{v_n(i)v_m^*(i)[ v_{m}(j)v_{n}^*(j) - u_{n}^*(j)u_{m}(j) ]}{\omega+i\eta-E_n+E_m} (-f(E_n)+f(E_m)) 
	\end{align}
	The remaining correlation functions (which we have not shown above) can be obtained by using the following symmetries,
	\begin{align}
		\chi_{ij}(j^x,\delta\Delta^\dagger) &= -\chi_{ji}(\delta\Delta,j^x) \\
		\chi_{ij}(j^x,\delta\Delta) &= -\chi_{ji}(\delta\Delta^\dagger,j^x) \\
		\chi_{ij}(\delta\Delta^\dagger,\delta\Delta^\dagger) &= \chi_{ji}(\delta\Delta,\delta\Delta) \\
		\chi_{ij}(\delta\Delta^\dagger,\delta n) &= \chi_{ji}(\delta n,\delta\Delta) \\
		\chi_{ij}(\delta n,\delta\Delta^\dagger) &= \chi_{ji}(\delta\Delta,\delta n). 
		\label{eq.G1}
	\end{align}
	Note that all the correlation functions reduce to the expressions provided in Ref.~\cite{bofan2021a} in the limit $T\rightarrow 0$. Now, we express, using Eq.~(\ref{eq.pha}), the different correlation functions in terms of the amplitude, phase and density fluctuations,
	\begin{align}
		\chi_{ij}(j^x,A)&=\frac{1}{\sqrt{2}} \left( \chi_{ij}(j^x,\delta\Delta) + \chi_{ij}(j^x,\delta\Delta^\dagger) \right)\\
		\chi_{ij}(A,j^x)&=\frac{1}{\sqrt{2}} \left( \chi_{ij}(\delta\Delta,j^x) + \chi_{ij}(\delta\Delta^\dagger,j^x) \right)\\
		\chi_{ij}(j^x,\Phi)&=\frac{i}{\sqrt{2}} \left( \chi_{ij}(j^x,\delta\Delta) - \chi_{ij}(j^x,\delta\Delta^\dagger) \right)\\
		\chi_{ij}(\Phi,j^x)&=\frac{i}{\sqrt{2}} \left( \chi_{ij}(\delta\Delta,j^x) - \chi_{ij}(\delta\Delta^\dagger,j^x) \right)\\
		\chi_{ij}(A,A)&=\frac{1}{2} \left[ \chi_{ij}(\delta \Delta,\delta \Delta) + \chi_{ij}(\delta \Delta,\delta \Delta^\dagger) + \chi_{ij}(\delta \Delta^\dagger,\delta \Delta) + \chi_{ij}(\delta \Delta^\dagger,\delta \Delta^\dagger) \right]\\
		\chi_{ij}(A,\Phi)&= \frac{i}{2} \left[ \chi_{ij}(\delta \Delta,\delta \Delta) - \chi_{ij}(\delta \Delta,\delta \Delta^\dagger) + \chi_{ij}(\delta \Delta^\dagger,\delta \Delta) - \chi_{ij}(\delta \Delta^\dagger,\delta \Delta^\dagger) \right]\\
		\chi_{ij}(A,\delta n)&= \frac{1}{\sqrt{2}} \left[ \chi_{ij}(\delta \Delta,\delta n) + \chi_{ij}(\delta \Delta^\dagger,\delta n) \right]\\
		\chi_{ij}(\Phi,A)&= \frac{i}{2} \left[ \chi_{ij}(\delta \Delta,\delta \Delta) + \chi_{ij}(\delta \Delta,\delta \Delta^\dagger) - \chi_{ij}(\delta \Delta^\dagger,\delta \Delta) - \chi_{ij}(\delta \Delta^\dagger,\delta \Delta^\dagger) \right]\\
		\chi_{ij}(\Phi,\Phi)&=\frac{i^2}{2} \left[ \chi_{ij}(\delta \Delta,\delta \Delta) - \chi_{ij}(\delta \Delta,\delta \Delta^\dagger) - \chi_{ij}(\delta \Delta^\dagger,\delta \Delta) + \chi_{ij}(\delta \Delta^\dagger,\delta \Delta^\dagger) \right]\\
		\chi_{ij}(\Phi,\delta n)&= \frac{i}{\sqrt{2}} \left[ \chi_{ij}(\delta \Delta,\delta n) - \chi_{ij}(\delta \Delta^\dagger,\delta n) \right]\\
		\chi_{ij}(\delta n,A)&= \frac{1}{\sqrt{2}} \left[ \chi_{ij}(\delta n,\delta \Delta) + \chi_{ij}(\delta n,\delta \Delta^\dagger) \right]\\
		\chi_{ij}(\delta n,\Phi)&= \frac{i}{\sqrt{2}} \left[ \chi_{ij}(\delta n,\delta \Delta) - \chi_{ij}(\delta n,\delta \Delta^\dagger) \right]
	\end{align}
	The above expressions are predictions for the different susceptibilities. Ultimately, our main focus is to calculate the current-current correlation functions that enter in the definition of the conductivity. However, it is necessary to compute the current-current correlations beyond the mean-field limit so our results are gauge invariant and can reproduce collective excitations that may impact the performance of the superinductor device. In order to proceed, we compute these corrections within the random phase approximation~\cite{anderson1958sc} that allows us to express them in terms of the susceptibilities above. Diagrammatically, these deviations correspond to vertex corrections to the mean-field bubble-diagrams representing the bare current-current correlation function.
	The full gauge invariant current-current correlation function is given by \cite{cea2014,seibold2017,bofan2021a}, 
	\begin{align}
		\chi_{ij}\left(j^x,j^x\right) = \chi^0_{ij}\left(j^x,j^x\right) + \Lambda_{ip}\mathbb{V}_{pl} \left(\mathbb{I}_{3N\times 3N}-\chi^B \mathbb{V}\right)^{-1}_{ls}\bar{\Lambda}_{sj}
		\label{eq.fullchi_SM}
	\end{align}
	where $\chi^0$ is the bare current-current correlation function (\ref{eq.A8}), and $\Lambda$ is a matrix whose entries are the susceptibilities above of the current with the amplitude fluctuations $A$, phase fluctuations $\Phi$ or charge density fluctuations $\delta n$,
	\begin{align}
		\Lambda&=\left(~\chi(j^x,A)~~\chi(j^x,\Phi)~~\chi(j^x,\delta n)~\right)\\
		\bar{\Lambda}&=\left(~\chi(A,j^x)~~\chi(\Phi,j^x)~~\chi(\delta n,j^x)~\right)^T.
	\end{align}
	$\chi^B$ is the bare mean-field susceptibility and $\mathbb V$ is the effective local interaction, defined by $3\times3$ matrices in the basis of fluctuations.
	\begin{equation}
		\chi^B = 
		\left(\begin{array}{ccc}
			\chi^{AA} & \chi^{A\Phi} & \chi^{A\delta n} \\
			\chi^{\Phi A} & \chi^{\Phi \Phi} & \chi^{\Phi \delta n} \\
			\chi^{\delta n A} & \chi^{\delta n\Phi} & \chi^{\delta n\delta n}
		\end{array}\right)
	\end{equation}
	and 
	\begin{equation}
		\mathbb{V} = 
		\left(\begin{array}{ccc}
			-|U| & 0 & 0 \\
			0 & -|U| & 0 \\
			0 & 0 & -|U|/2
		\end{array}\right)
	\end{equation}
	For example, in Eq.~\eqref{eq.fullchi_SM}, $\chi^B(A,A)=\chi^{AA}$, $\mathbb{V}^A=-|U|\mathbb{I_{N\times N}}$. Note that all of these correlation function are $N\times N$ matrices in real space.
	
	The complex optical conductivity $\sigma(\omega) = \sigma_1(\omega) - i\sigma_2(\omega)$ is obtained ~\cite{cea2014,cea2015,seibold2015,seibold2017} directly from the susceptibility $\chi(\omega)$,
	\begin{align}
		\sigma_1(\omega) &= \pi D_s\delta(\omega) + e^2{\rm Im} \frac{\chi(\omega)}{\omega} \label{eq.sigma1}\\
		\sigma_2(\omega) &= e^2 \frac{\langle -k_x \rangle + {\rm Re} \chi(\omega)}{\omega} \label{eq.sigma2}
	\end{align} 
	where $\rm Im$ stands for the imaginary part of $\chi(\omega) = 1/N\sum_{ij}\chi_{ij}(j^x,j^x)$, while $\rm Re$ is the real part. $e$ is the elementary charge, $D_s = e^2 [\langle -k_x \rangle + {\rm Re} \chi(\omega=0)]$ is the superfluid stiffness, and $\langle -k_x \rangle = {4t\over N}\sum_{i}\sum_{n}[u_n(i)u_n(i+x) f(E_n) + v_n(i)v_n(i+x)(1-f(E_n))]$ is the kinetic energy along the $x$ direction.
	
	The quality factor of a transmission line is defined as \cite{devisser2014} $Q = {\alpha\over 2\beta}$, where $\alpha$ and $\beta$ are the real and imaginary parts of the propagation constant $\gamma$. The superinductor which absorbs electromagnetic radiation can be treated as a lossy media. Therefore, the complex propagation constant of the plane wave can be written as \cite{yassin1995electromagnetic,cheng1989field} 
	\begin{align}
		\gamma = \sqrt{i\omega \mu(\sigma+i\omega\epsilon)} = \alpha + i\beta,
	\end{align}  
	where $\epsilon$ and $\mu$ are the permittivity and permeability of the media, respectively. With the assumption $\sigma_2 \gg \sigma_1$, one can easily obtain
	\begin{equation}
		\begin{aligned}
			Q &={\alpha \over 2\beta} \\
			&= {\cos\lt({\arctan{\sigma_1\over \sigma_2 - 2\pi f\epsilon}\over2}\rt)  \over 2\ {\sin\lt({\arctan{\sigma_1\over \sigma_2 - 2\pi f\epsilon}\over2}\rt)}} \\
			&\approx {\sigma_2 \over \sigma_1 }. 
		\end{aligned} \label{eq.Q_planar}
	\end{equation}
	
	Likewise, the kinetic inductance is also expressed in terms of the complex conductivity $L_{k/\square} = {1\over 2\pi f \sigma_2}$ \cite{kamenov2020granular,tinkham2004}.

	\section{Estimation of finite size effects}
\begin{figure}[!htbp]
		\begin{center}
			\subfigure[]{\label{fig.gap_size} 
				\includegraphics[width=5.2cm]{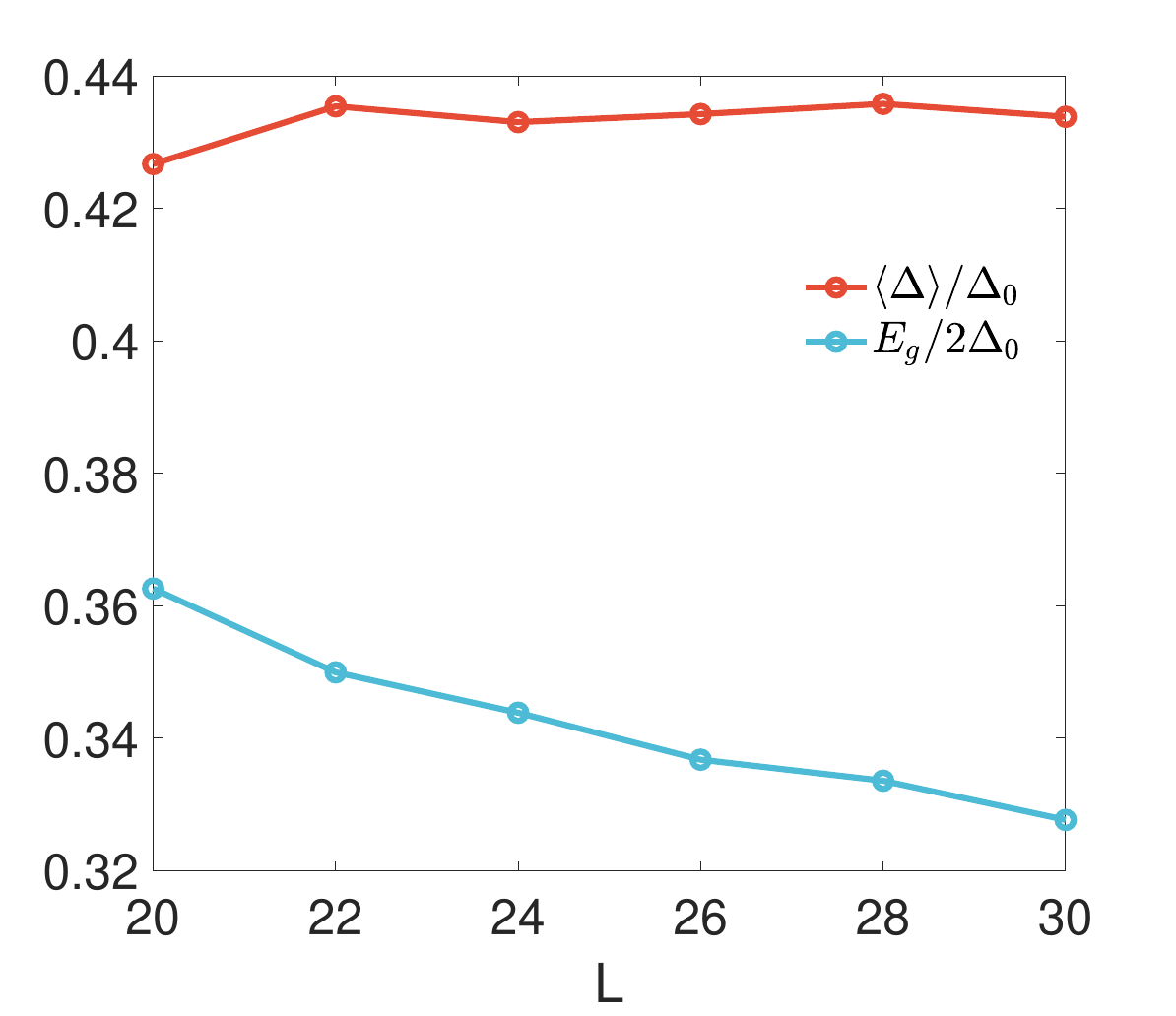}}
			\subfigure[]{\label{fig.sizeQ} 
				\includegraphics[width=5.2cm]{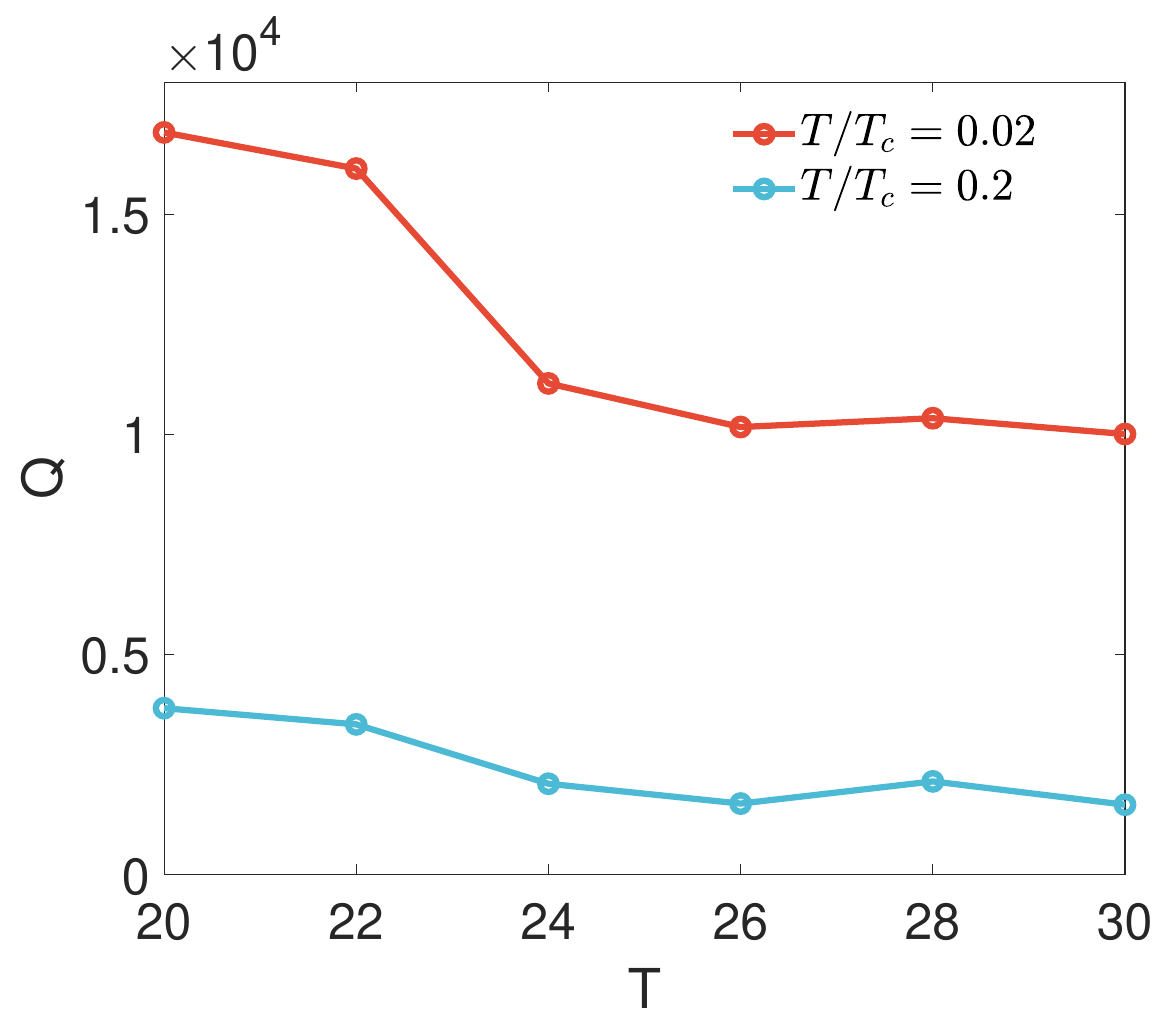}}
			\subfigure[]{\label{fig.sizeLk} 
				\includegraphics[width=5.2cm]{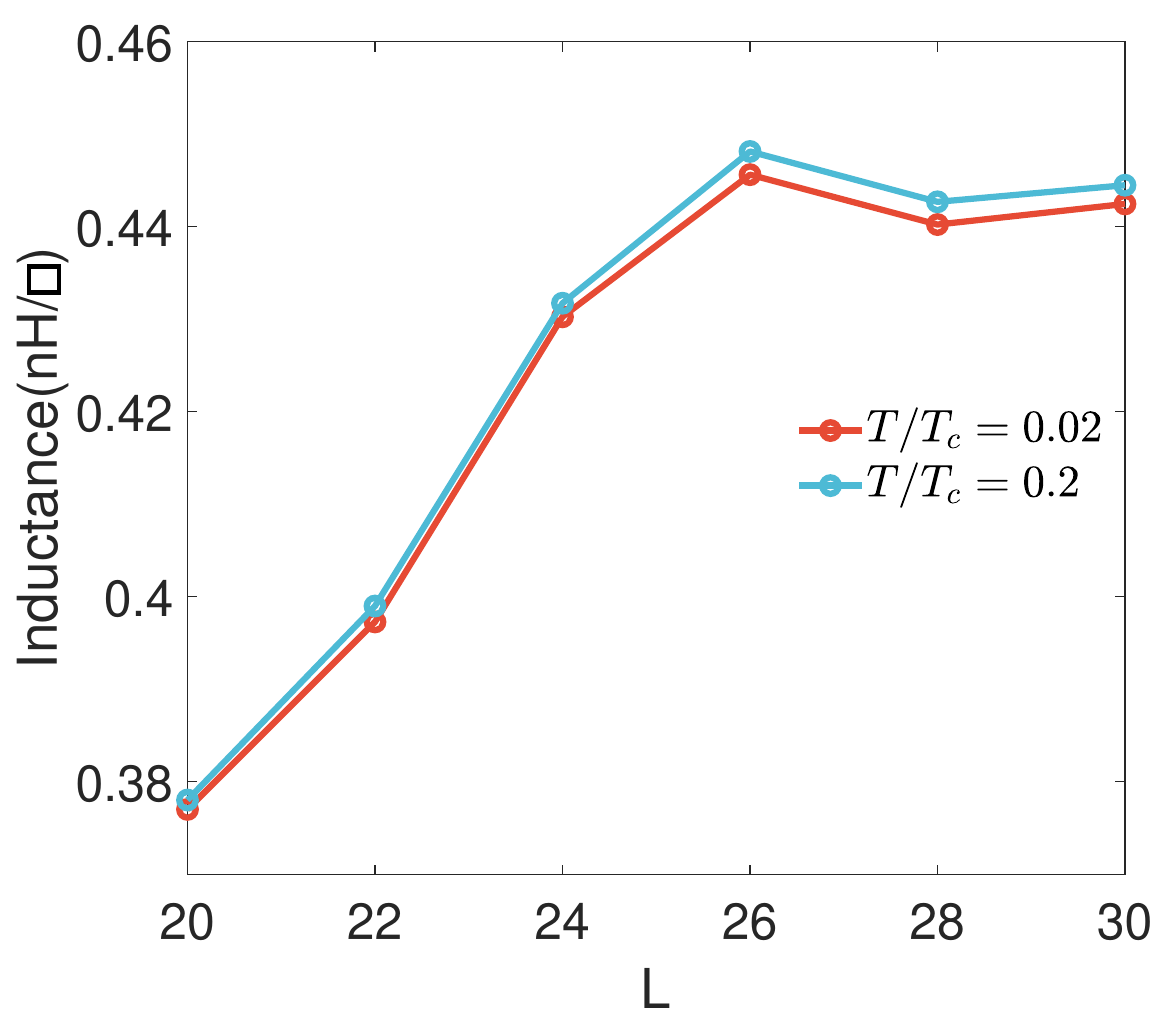}}
			\caption{\subref{fig.gap_size}. The spatial averaged order parameter $\langle \Delta\rangle$ and the 
				spectral gap $E_g$, normalized by factors of $\Delta_0$. It is evident that $\langle \Delta\rangle$ changes only slightly with size. \subref{fig.sizeQ} The Quality factor $Q$ and \subref{fig.sizeLk} kinetic inductance $L_{k/\square}$. The disorder is $V=1.5$, and the frequency is $f = 0.112\Delta_0$. The other parameters are $U = -1, \langle n \rangle = 0.875$. 
				We do not observe any substantial size dependence in $Q$ or $L_{k/\square}$, when $L\geq 26 $ at $V=1.5$. For stronger coupling or disorder strength, finite size effects will be even smaller.}\label{Fig:sizeQs}
		\end{center}
	\end{figure}

	Due to the limitation of our computational resources, the maximum size we could reach is $N=L\times L = 32\times32$. For this size, there are strong size effects for $V = 0$, no disorder, because the coherence length $\xi$ is much larger than the system size in the weak-coupling limit of interest. We expect that as $V$ increases, the coherence length decreases and finite size effects will become negligible.  
	In Fig.~\ref{Fig:sizeQs}, we show the size dependence of the spatial averaged order parameter $\langle \Delta\rangle$ and 
	spectral gap $E_{g}$, the quality factor $Q$, and the kinetic inductance $L_{k/\square}$ for $V =1.5$. The size dependence is quite weak when $L\geq 26$, so finite size effects are not important in the region close to the transition $V\gtrsim 1.5$ which is the main focus of the paper.

	\newpage

	\section{The Quality factor computed from the sample-averaged conductivity}
	In the main text, the quality factor $Q$ was computed from the sample-averaged conductivity, namely, we compute the conductivity for each disorder realization, then we perform the average over all disorder realizations. Finally, the quality factor was obtained from this averaged quantity.  
	This is justified because 
	our system size is small compared to the experimental one. Note that assuming a lattice spacing of $0.35$nm, our typical lattice size is $~10$ nm $\times 10$ nm.
	
\begin{figure}[!htbp]
		\begin{center}
			\includegraphics[width=7.5cm]{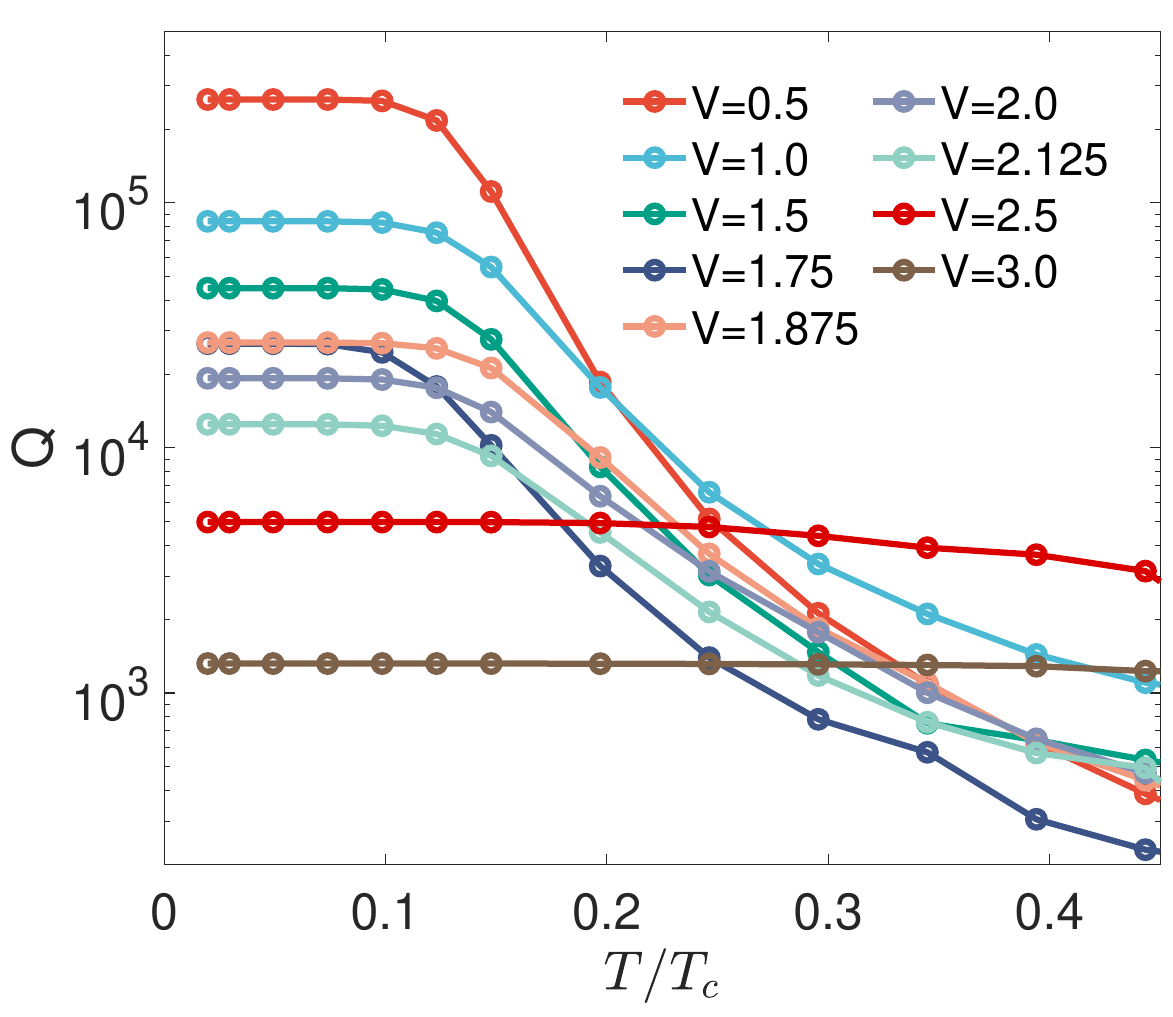}
			\includegraphics[width=7.5cm]{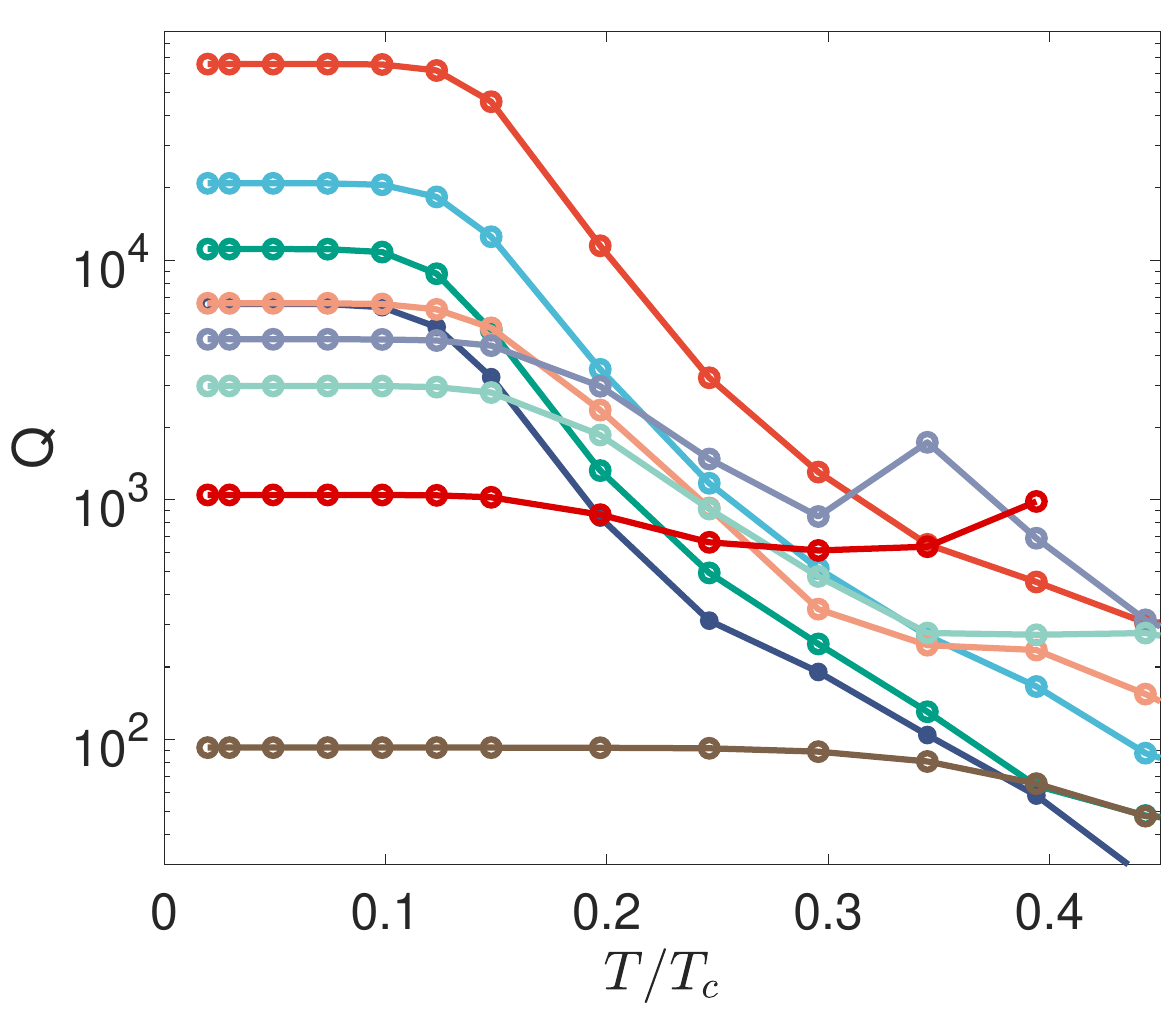}
			\caption{Quality factor as a function of temperature for different disorder strength $V$ with sample-averaged conductivity. The frequencies are $f = 0.028\Delta_0$ (left panel) and $f = 0.112\Delta_0$ (right panel). Other parameters are $N=28\times28, U = -1, \langle n \rangle = 0.875$. 
			}\label{Fig:Qc_U1}
		\end{center}
	\end{figure}

	For a single configuration of the conductivity, since the size is not large enough, we might get some spurious sub-gap excitation peaks which would lead to huge sample-sample variations in the quality factor.
	In experiments, since the sample size is much larger, it can access all possible excitations within a single sample.
	Therefore, by first performing the average of the conductivity, we reduce significantly the statistical error. In Fig.~\ref{Fig:Qc_U1}, we depict the quality factor for all disorder values we have studied. As in the main text, the sample-average of the conductivity was carried out first,
	We shall see in next section that these results are consistent with those obtained by computing the quality factor for each disorder realization and then performing the ensemble average over all disorder realizations. 

	\section{The sample averaged Quality factor and Kinetic Inductance}
\begin{figure}[!htbp]
		\begin{center}
			\subfigure[]{\label{fig.sizeQ_a} 
				\includegraphics[width=7.5cm]{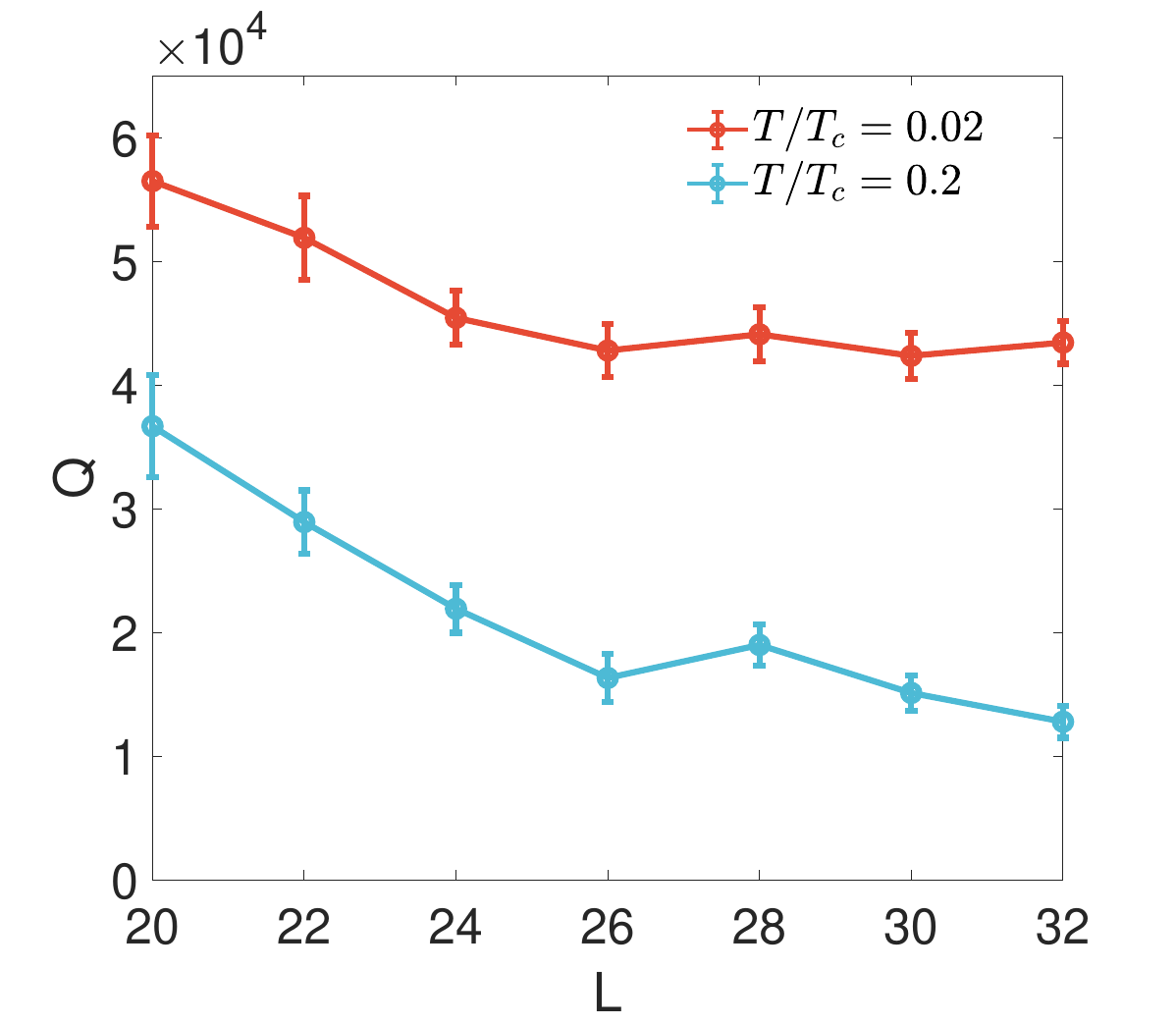}}
			\subfigure[]{\label{fig.sizeQ_b} 
				\includegraphics[width=7.7cm]{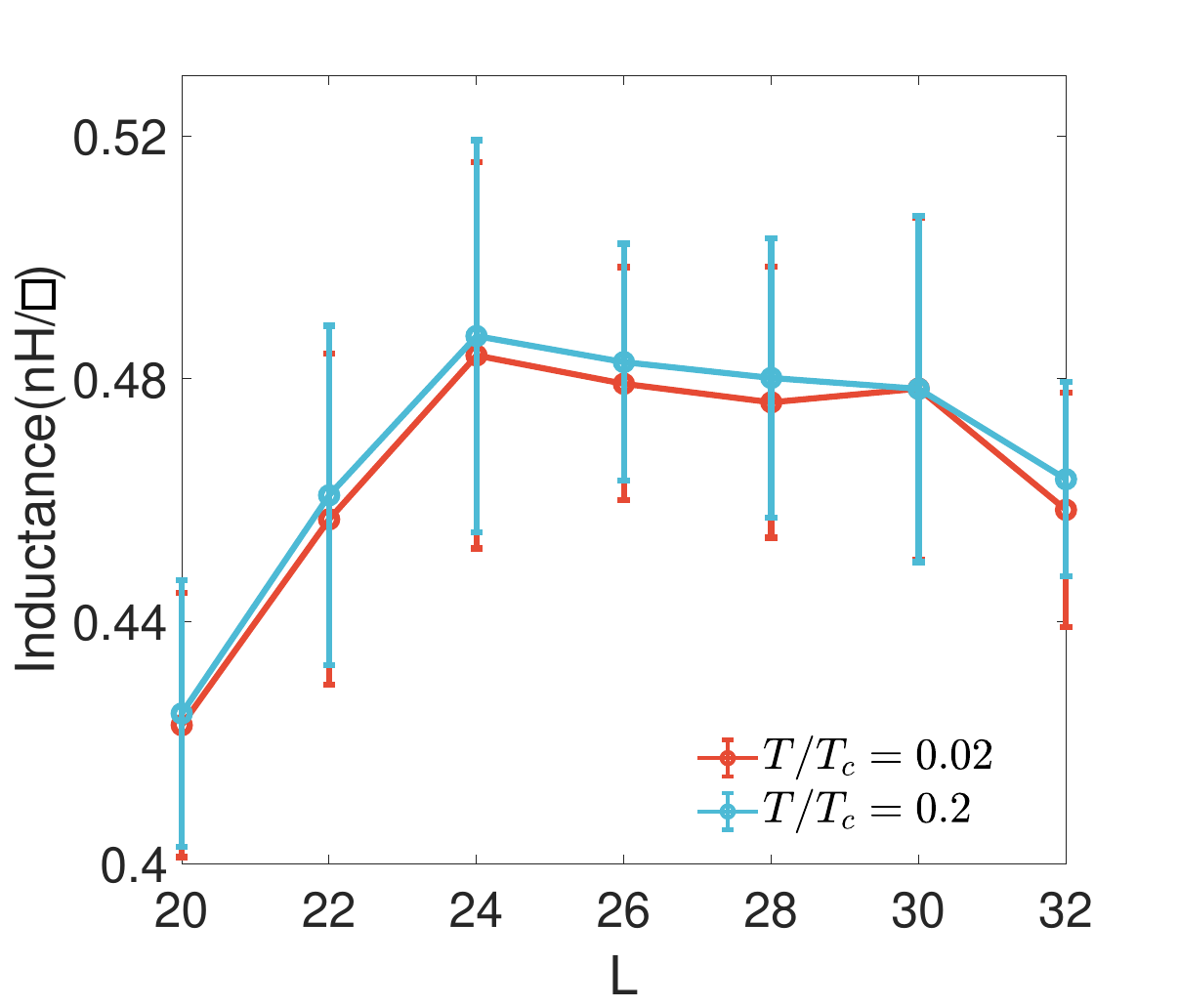}}\\
			\subfigure[]{\label{fig.sizeQ_c} 
				\includegraphics[width=7.5cm]{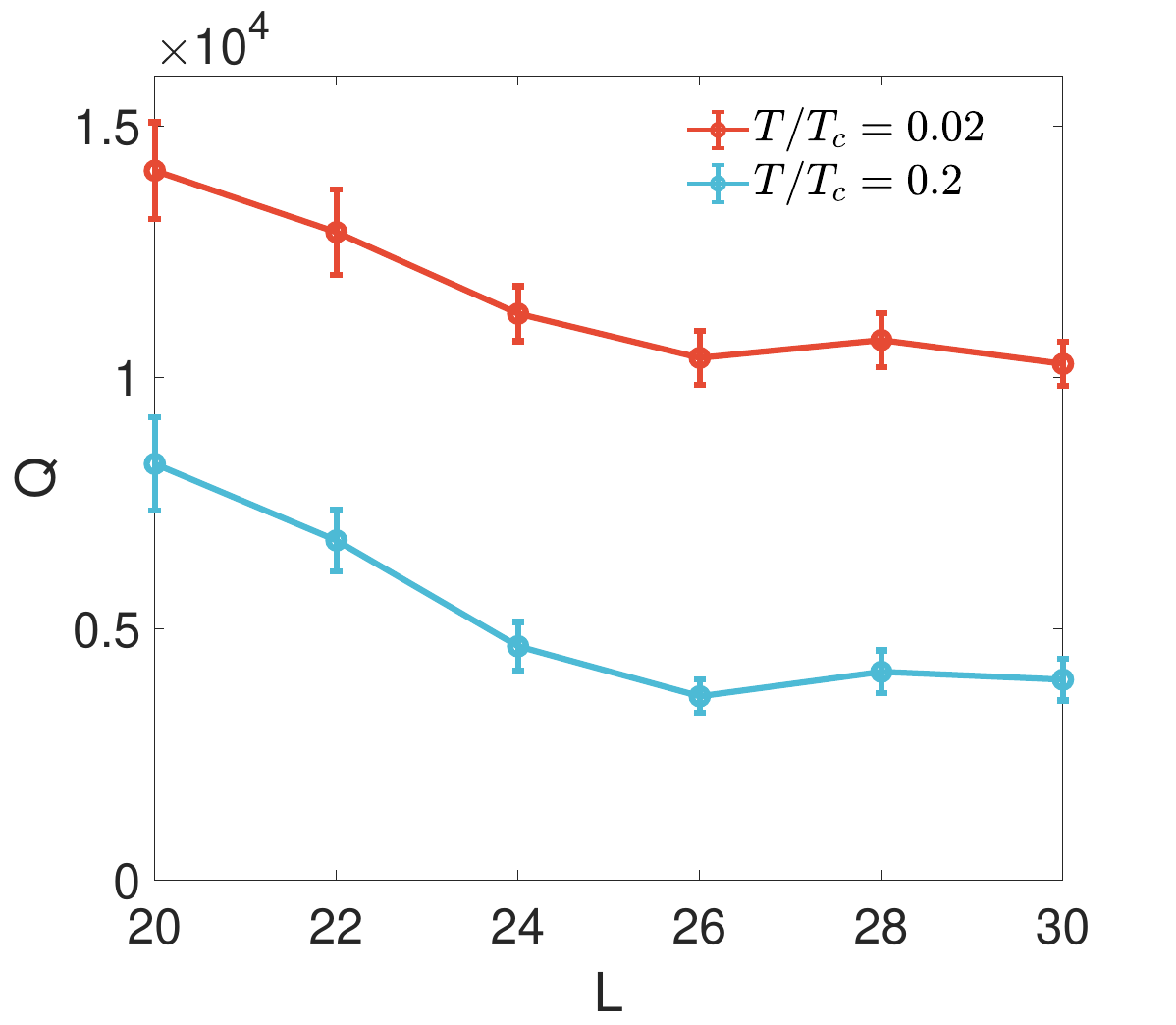}}
			\subfigure[]{\label{fig.sizeQ_d} 
				\includegraphics[width=7.7cm]{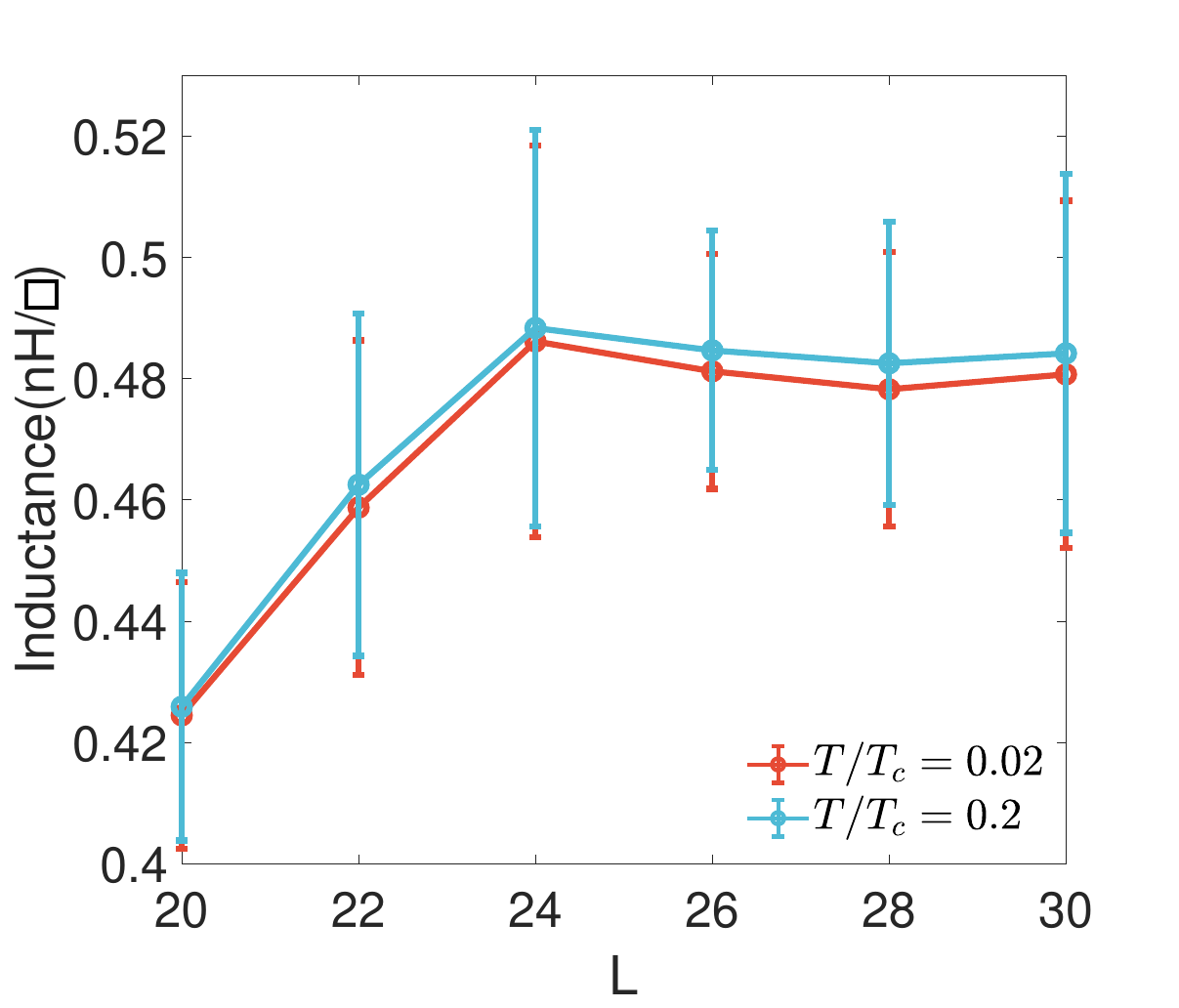}}
			\caption{The Quality factor $Q$ (left) and kinetic inductance $L_{k/\square}$ (right) as a function of system size. The disorder is $V=1.5$ and the frequencies are $f = 0.028\Delta_0$ (top row) and $f = 0.112\Delta_0$ (bottom row). 
				Other parameters are the same as those of Fig.~\ref{Fig:sizeQs}. Here, we have done the sample averaging in the final step while calculating $Q$ and $L_{k/\square}$. Notice that the results are consistent with Fig.~\ref{Fig:sizeQs}. }\label{Fig:sizeQ}
		\end{center}
	\end{figure}

	In this section, we present results of the sample-averaged quality factor $Q$ and kinetic inductance $L_{k/\square}$. By sample-average, we mean that, unlike the previous two sections, we compute $Q$ and $L_{k/\square}$ for each disorder realization and, after that, we perform the ensemble average.  
	In Fig.~\ref{Fig:sizeQ}, we depict the size dependence of $Q$ and $L_{k/\square}$ as a function of the system size. 
	These results are very similar to those of Fig.~\ref{Fig:sizeQs} where $Q$ and $L_{k/\square}$ were computed by first carrying out the disorder average of the conductivity.
	
\begin{figure}[!htbp]
		\begin{center}
			\subfigure[]{\label{fig.Qs_a} 
				\includegraphics[width=7.5cm]{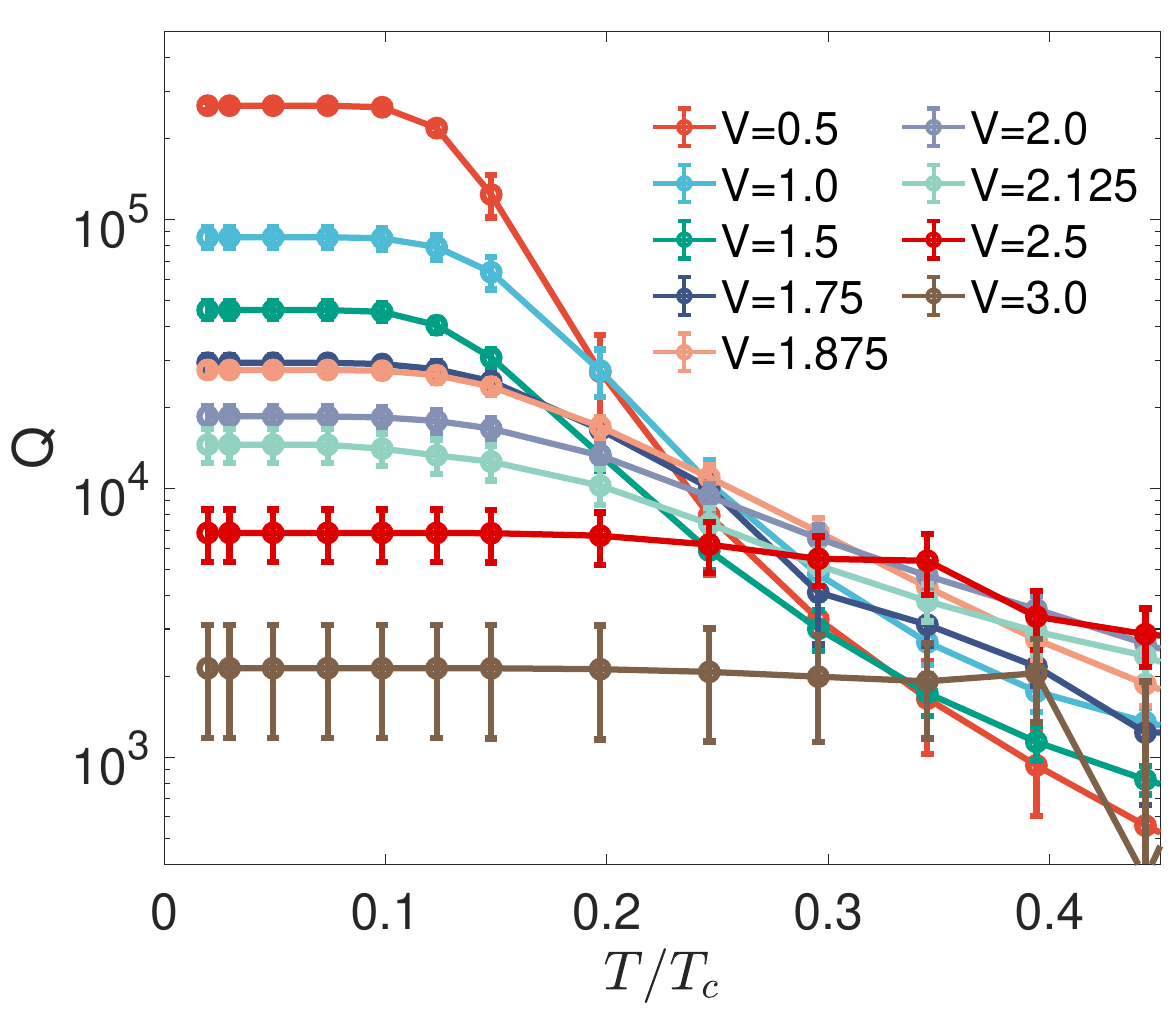}}
			\subfigure[]{\label{fig.Qs_b} 
				\includegraphics[width=7.5cm]{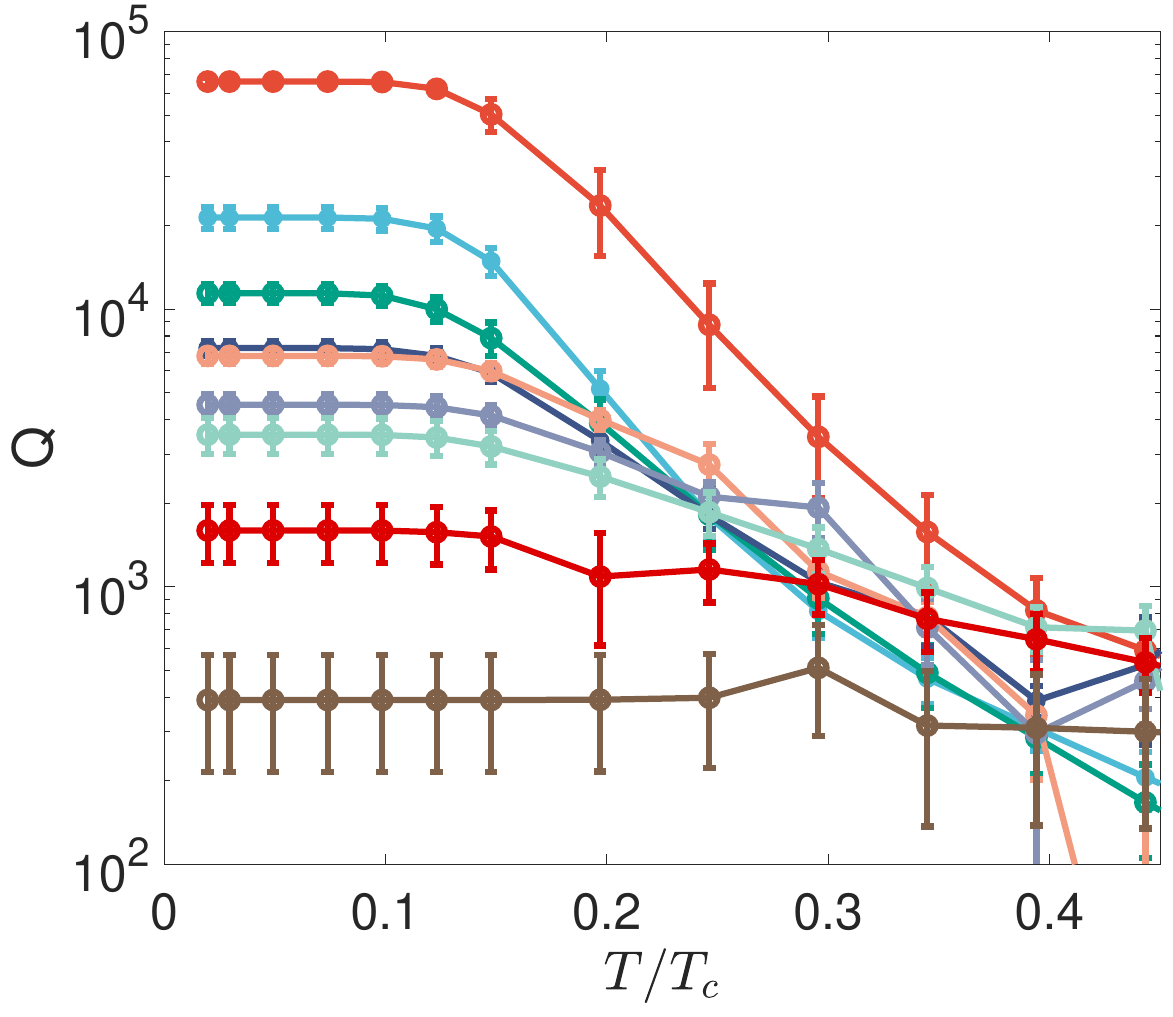}}\\
			\subfigure[]{\label{fig.Qs_c} 
				\includegraphics[width=7.6cm]{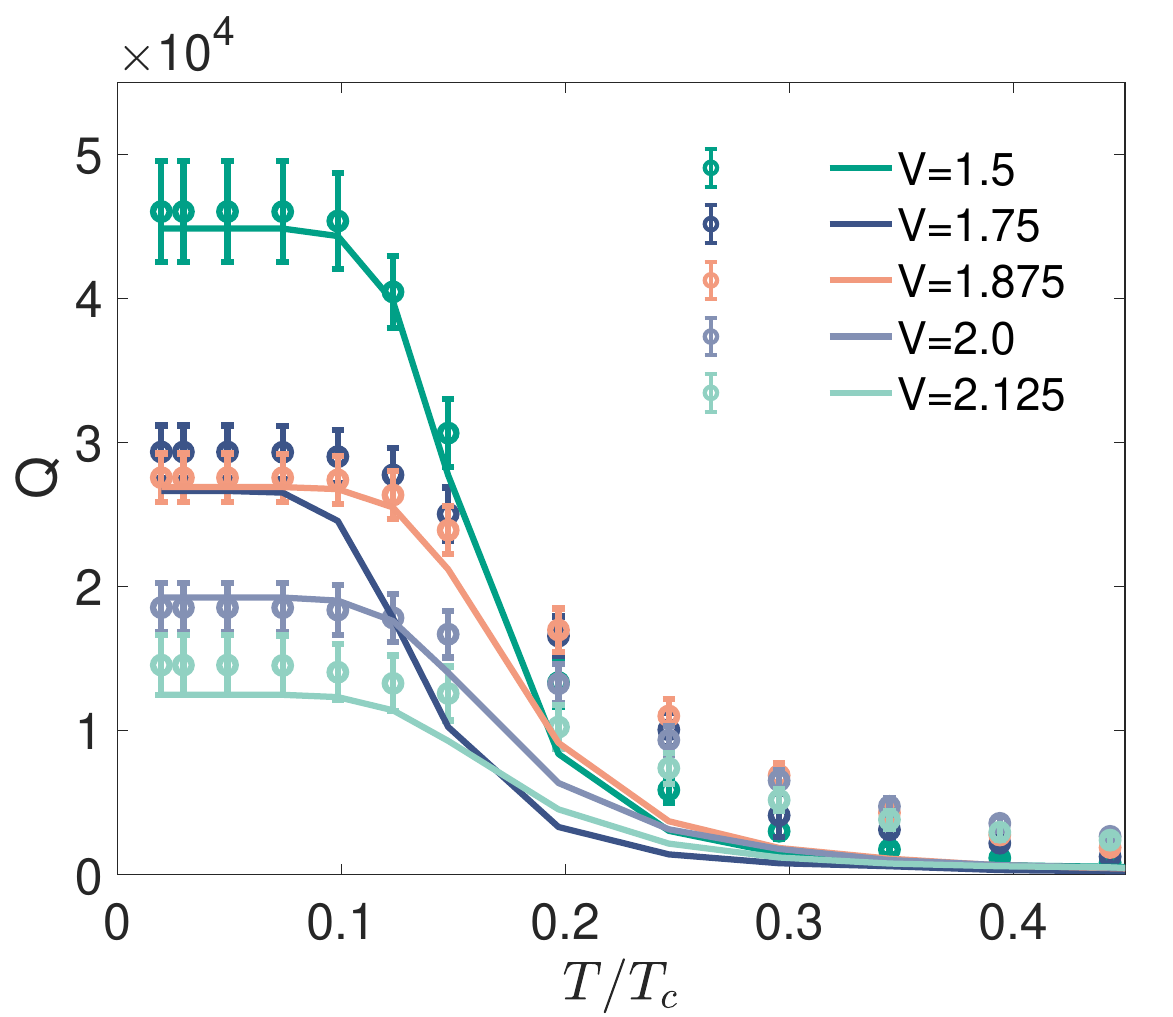}}
			\subfigure[]{\label{fig.Qs_d} 
				\includegraphics[width=7.6cm]{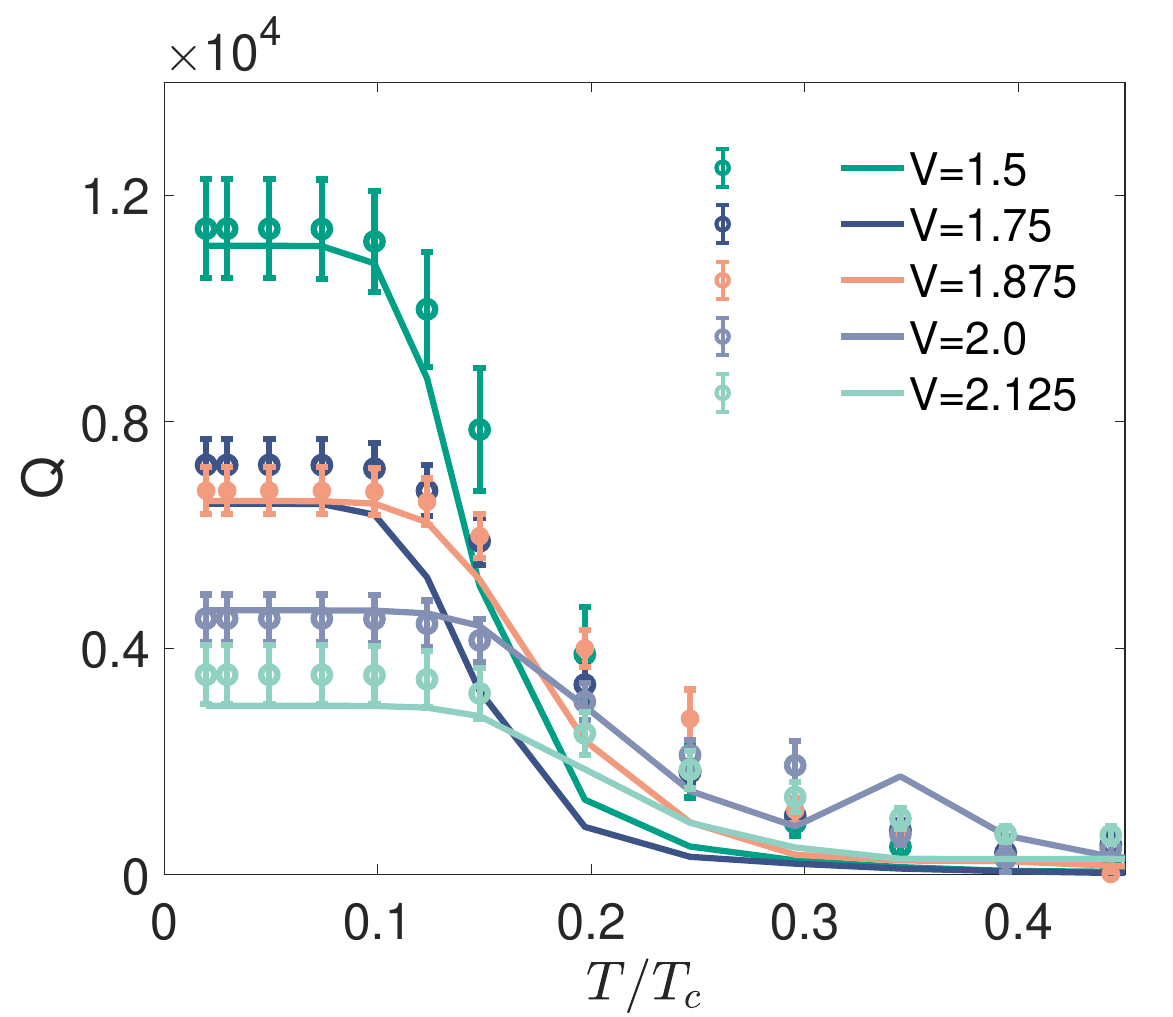}}
			\caption{Quality factor as a function of temperature for different disorder strength for a lattice $N = 28\times28, U = -1, \langle n \rangle = 0.875$. The frequency for \subref{fig.Qs_a} is $f = 0.028\Delta_0$, and for \subref{fig.Qs_b} is $f = 0.112\Delta_0$. The lower panel includes the same plots with above but in linear scale, and only around the critical disorder. The sample averaged quality factors are presented by the circles with error bar, and the solid lines are the corresponding results based on sample averaged conductivity results. The deviations between two methods are small. In the clean limits, we have $\Delta_0 = 0.0357t,T_{c} \approx 0.02t$ based on the BCS assumption. 
			}\label{Fig:Qs_U1}
		\end{center}
	\end{figure}

\begin{figure}[!htbp]
		\begin{center}
			\includegraphics[width=7.5cm]{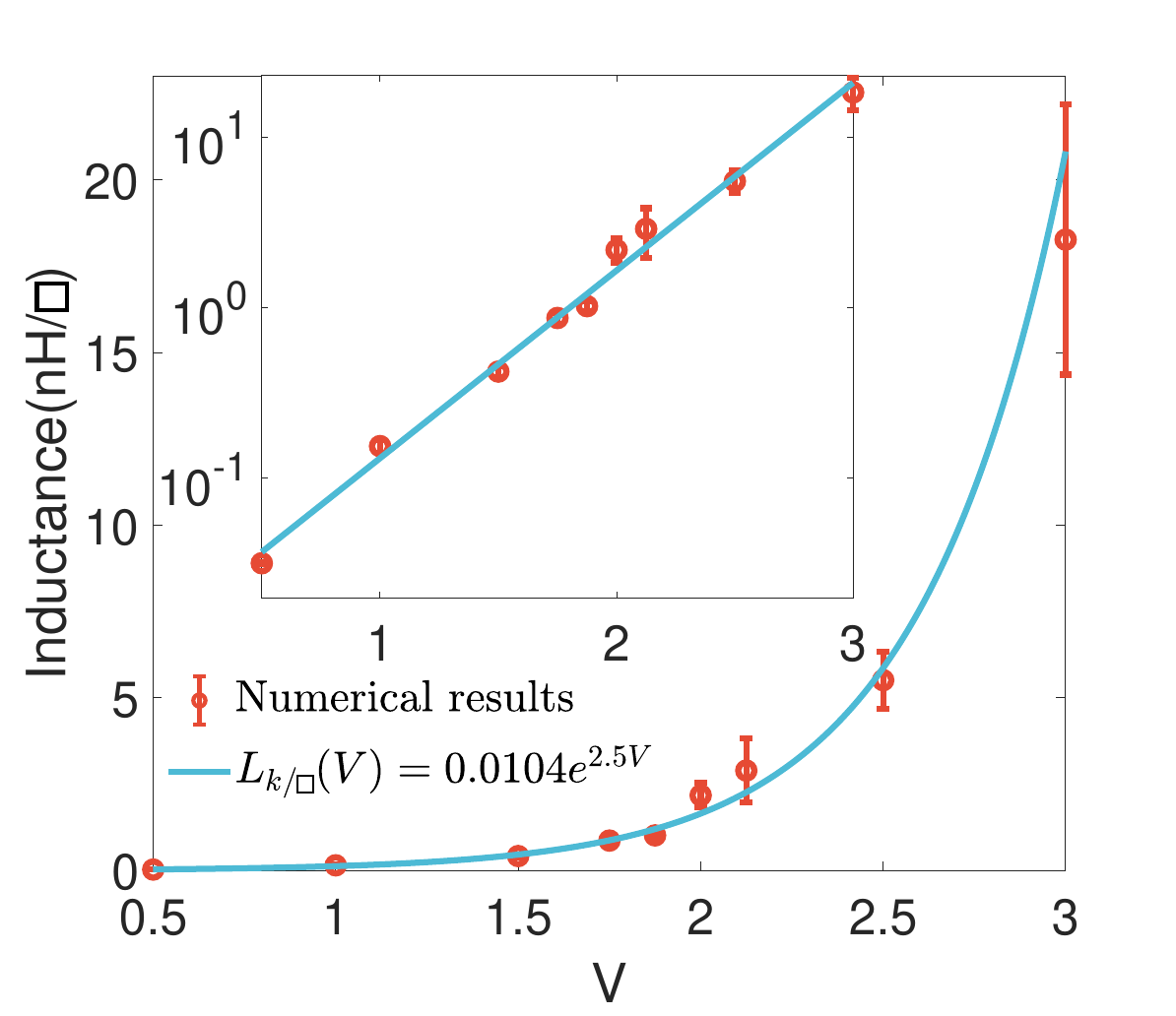}
			\caption{ The kinetic inductance $L_k$ as a function of disorder $V$. The numerical results fits well with an exponential increase of $L_k$ with disorder. The fitting exponent is $2.5$, which is consistent with the result $2.3$ in the main text. The other parameters are $N=28\times28, U = -1, \langle n \rangle = 0.875$.}\label{Fig:L_ks}
		\end{center}
	\end{figure}

	Next we compute $Q$ as a function of the temperature for different values of the disorder strength. If there are enough disorder realizations,  the quality factor $Q$ computed by averaging over disorder, see Fig.~\ref{Fig:Qs_U1}, is almost identical than the one performing first the average over the conductivity and then computing the  quality factor, see Fig.~\ref{Fig:Qc_U1}. The exponential increase of the kinetic inductance  $L_{k/\square}$, see Fig.~\ref{Fig:L_ks}, is also observed by this approach with results consistent with those presented in the main text.

	\section{The Quality factor for a cavity resonator}
	Based on the prediction of high frequency dissipation in a superconducting cavity, the quality factor is given by~\cite{tinkham2004} 
	\begin{equation}
		Q = {f\over 2c}{{\mathcal V}\over S}{\pi({\sigma_2\over 4\pi \epsilon})^2\over {\sigma_1\over 4\pi \epsilon}\sqrt{2\pi f(|{\sigma\over 4\pi \epsilon}|+{\sigma_2\over 4\pi \epsilon})}}
		\approx {\pi({\sigma_2\over 4\pi \epsilon})^2\over 2{\sigma_1\over 4\pi \epsilon}\sqrt{2\pi f(|{\sigma\over 4\pi \epsilon}|+{\sigma_2\over 4\pi \epsilon})}}, \label{eq.Q_cavity}
	\end{equation}
	where $c$ is the speed of light, ${\mathcal V}$ is the volume of the superconductor cavity and $S$ is the corresponding surface area. For the sake of clarity, we use the conventional symbol $f$ for the frequency to replace $\omega$. Here, we are assuming that the superconductor cavity is operating at the lowest modes, so that the typical linear dimension of the cavity $l={{\mathcal V}\over S}$ is of the order of the radiation wavelength $\lambda = {c/f}$.
	
	In the limit of $\sigma_2 \gg \sigma_1$, Eq.\eqref{eq.Q_cavity} can be approximated by
	\begin{equation}
		Q \approx \sqrt{\pi\over 4}\sqrt{\sigma_2\over4\pi f \epsilon} {\sigma_2\over \sigma_1} \label{eq.Q_cavity_approx}.
	\end{equation}
	It has been shown \cite{devisser2014} that in a microwave resonator, the working frequency of the superinductor, which is near the resonate frequency $f$, is proportional to $\sigma_2$. Therefore, comparing Eqs.~\eqref{eq.Q_planar} and \eqref{eq.Q_cavity_approx}, the quality factor in the cavity case should qualitatively agree with the planar case, up to a  prefactor $\sqrt{\pi\over 4}\sqrt{\sigma_2\over4\pi f \epsilon}$ which only depends weakly on disorder and temperature. 
	
\begin{figure}[!htbp]
		\begin{center}
			\subfigure[]{\label{fig.Qc_a} 
				\includegraphics[width=7.5cm]{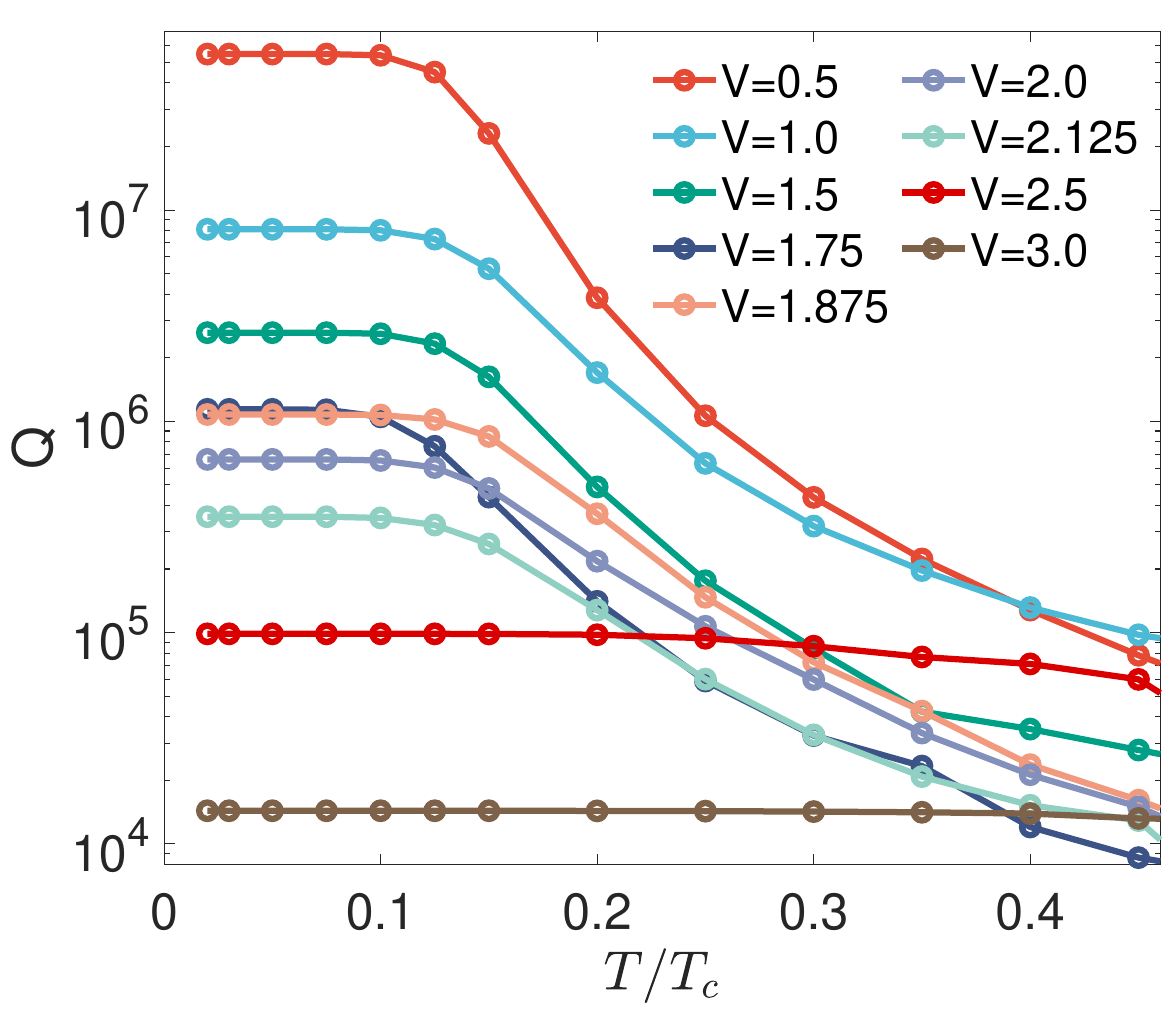}}
			\subfigure[]{\label{fig.Qc_b} 
				\includegraphics[width=7.5cm]{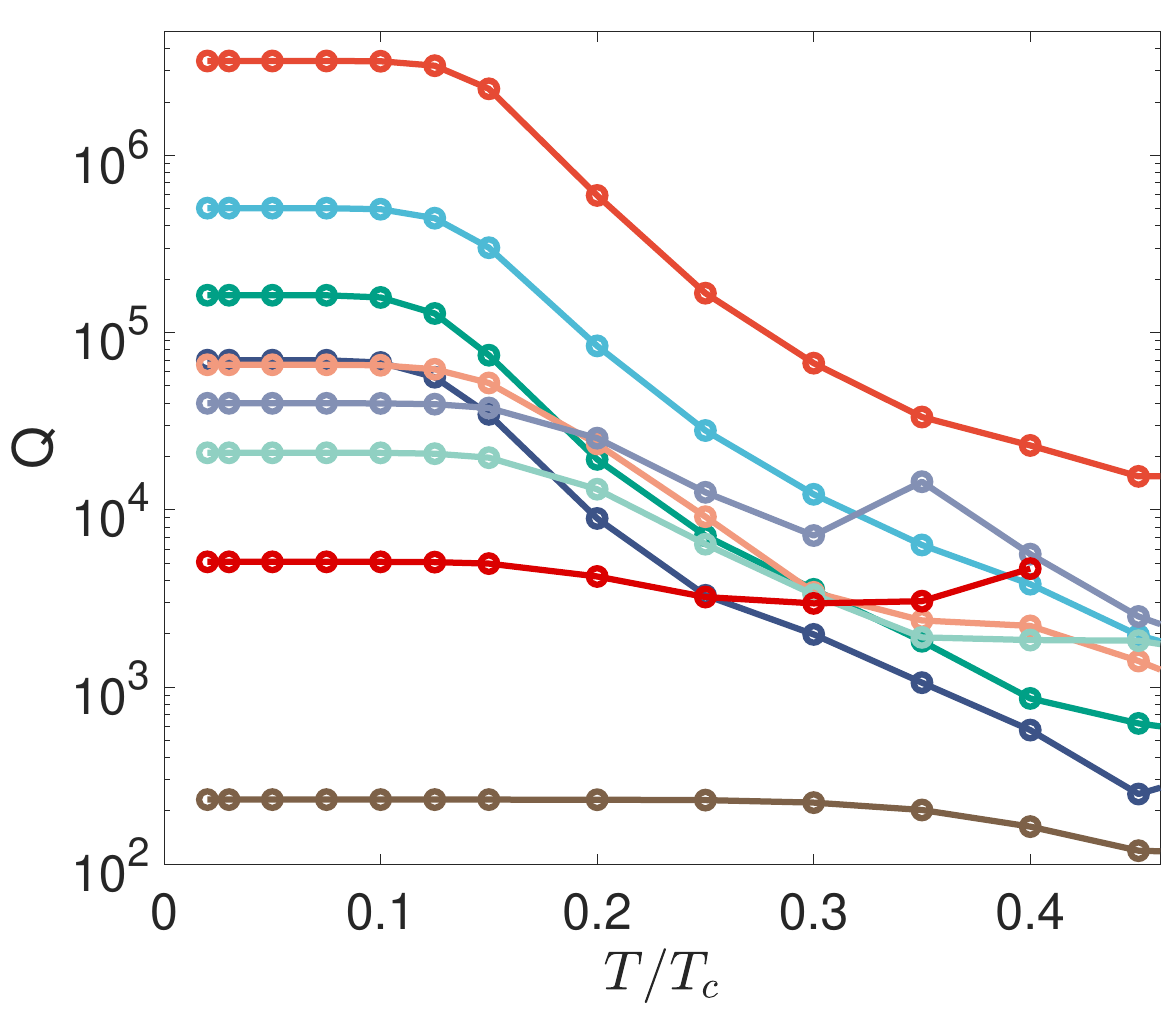}} \\
			\subfigure[]{\label{fig.Qc_c} 
				\includegraphics[width=7.5cm]{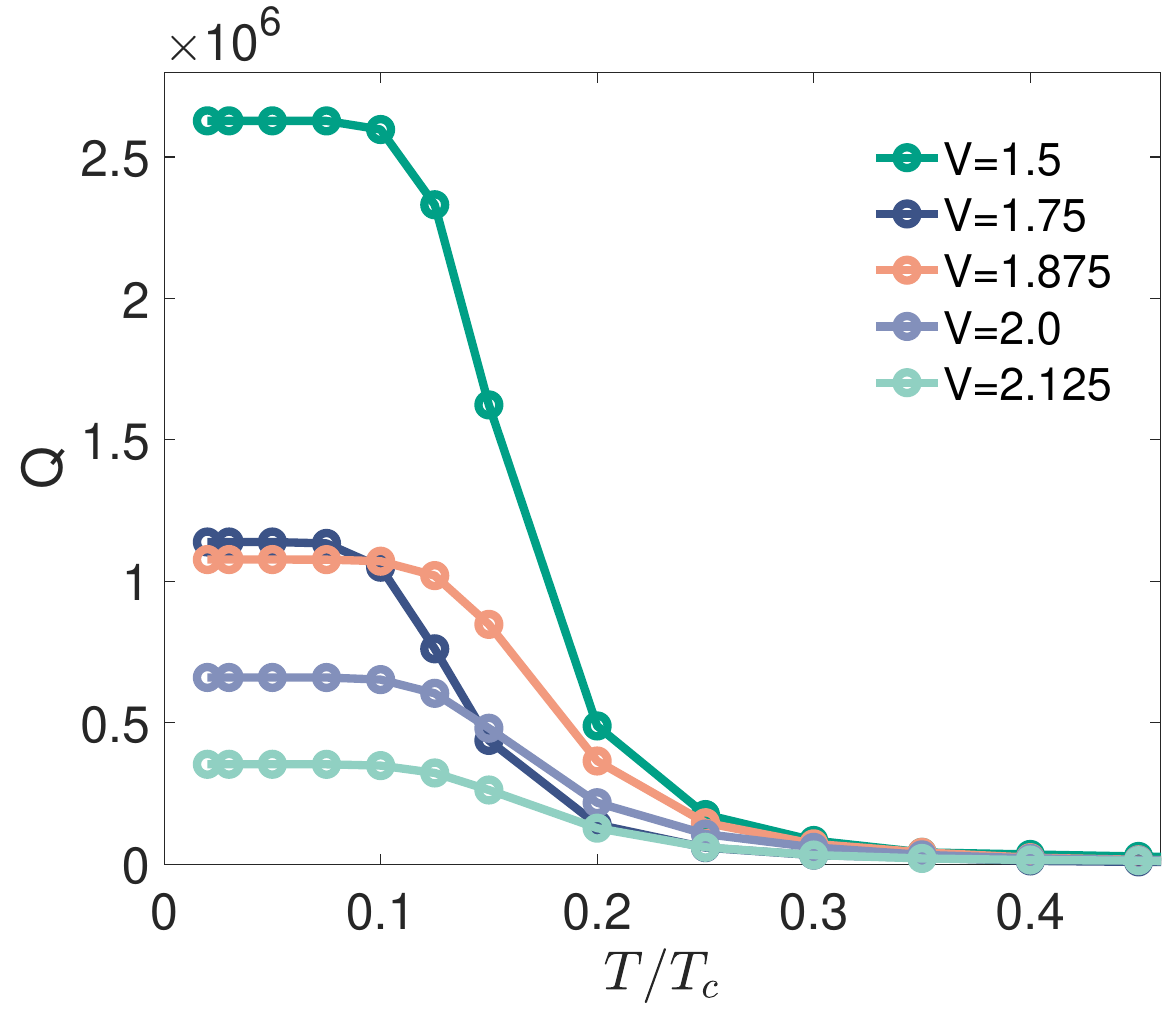}}
			\subfigure[]{\label{fig.Qc_d} 
				\includegraphics[width=7.5cm]{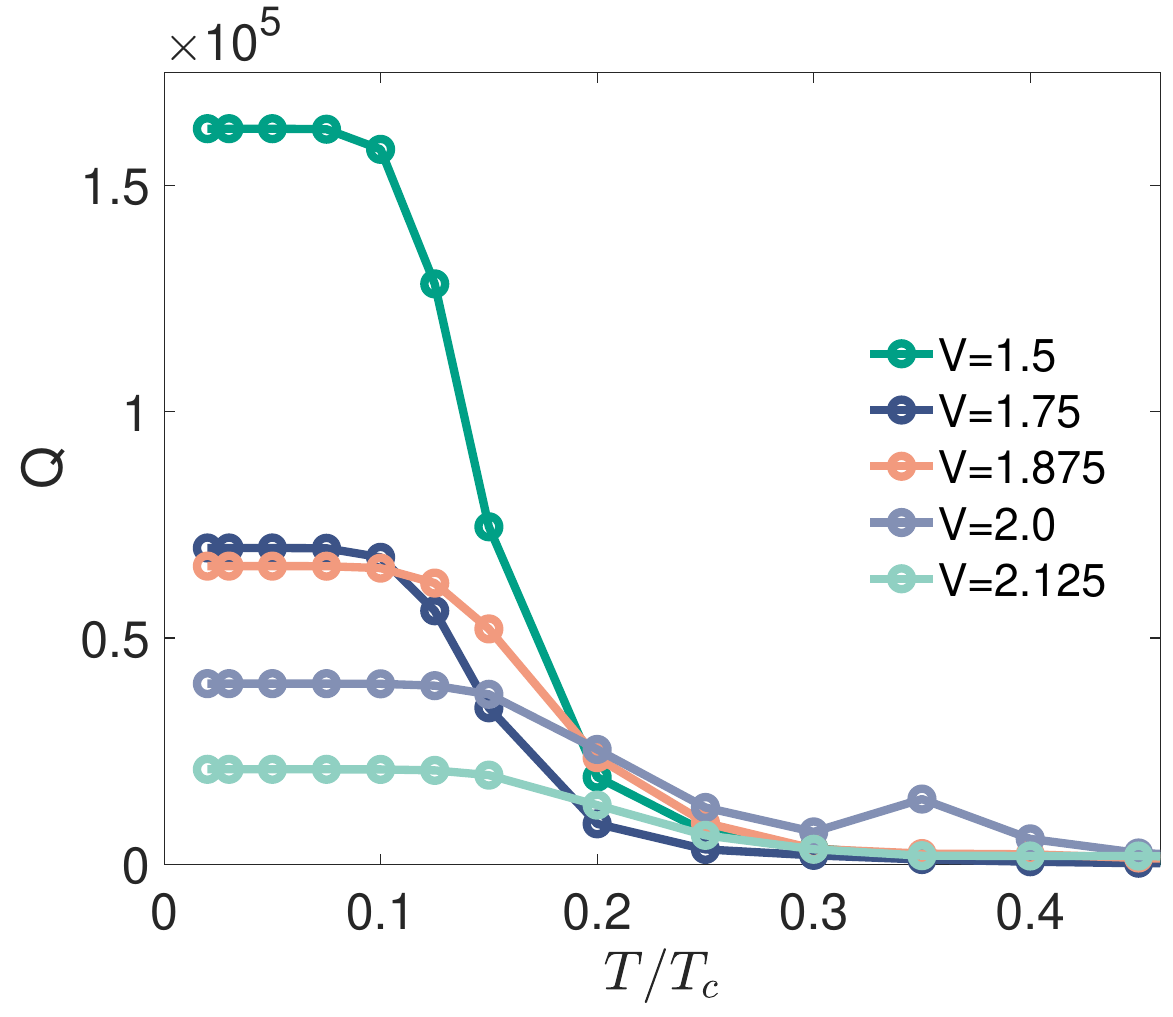}}
			\caption{
				Quality factor $Q$ as a function of temperature for different disorder strength for a lattice $N = 28\times28, U = -1, \langle n \rangle = 0.875$. The frequency for \subref{fig.Qc_a} is $f = 0.028\Delta_0$, and for \subref{fig.Qc_b} is $f = 0.112\Delta_0$. The lower panel contains the same information but in linear scale, and only around the critical disorder. The quality factor are calculated by using Eq.\eqref{eq.Q_cavity} based on sample averaged conductivity results. 
			}\label{Fig:Q_cavity}
		\end{center}
	\end{figure}

\begin{figure}[!htbp]
		\begin{center}
			\includegraphics[width=7.5cm]{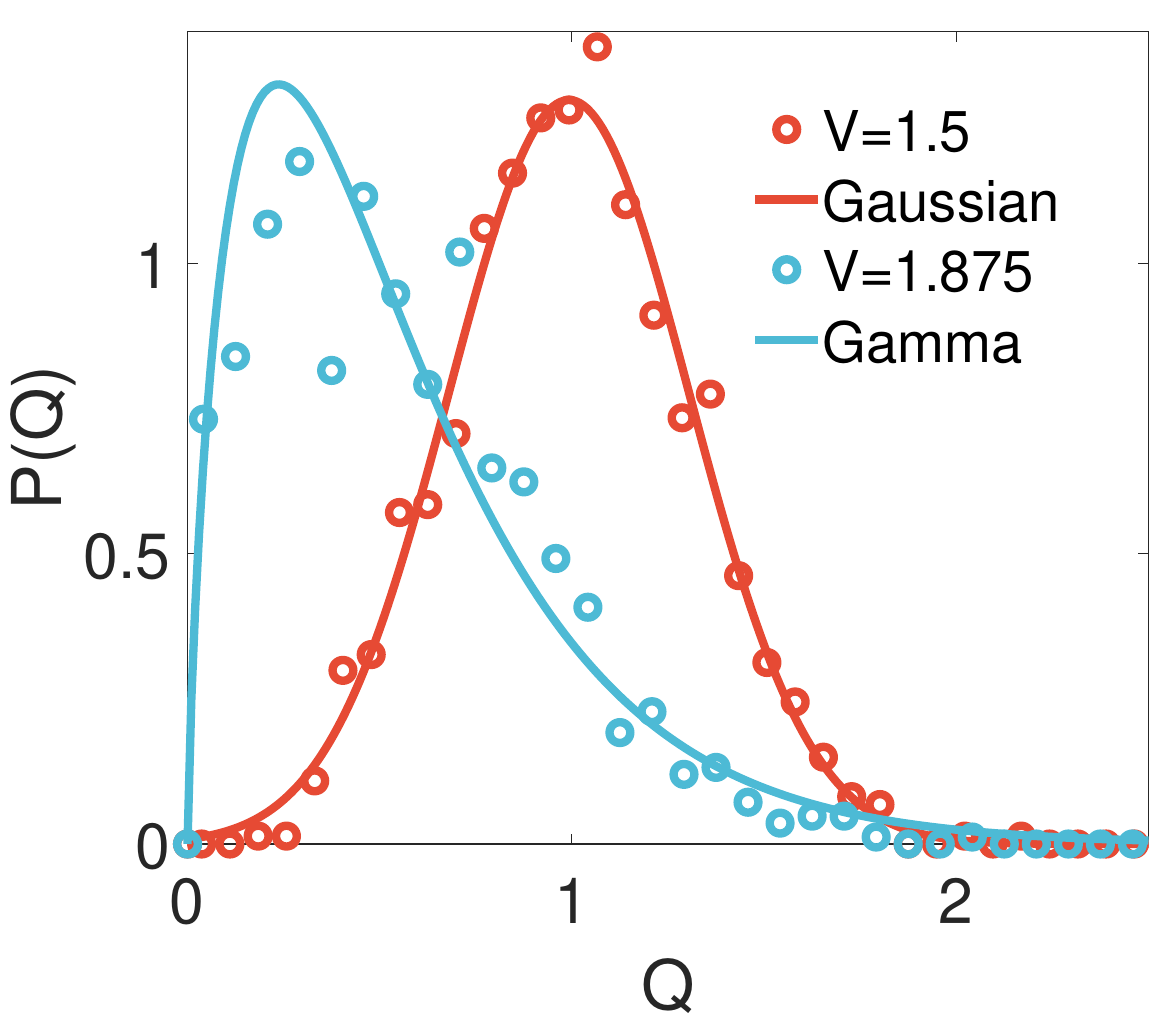}
			\caption{Probability distribution of $Q$ from $1000$ disorder realizations, normalized by the average of $Q (V=1.5) \sim 2.5\times10^6$ for $f =  0.028\Delta_0\sim 2.5$ GHz, and temperature $T/T_c = 0.05$. When $V=1.5$, the distribution of $Q$ fits well with a Gaussian distribution, red line. However, very close to the transition
				$V \approx 1.9\approx V_c$, the distribution, asymmetric and with broad tails, is similar to a Gamma distribution (blue line). The results are similar to those of the planar case presented in the main text.
			}\label{Fig:Pro_Q_cavity}
		\end{center}
	\end{figure}

	The quality factor, see Figs.~\ref{Fig:Q_cavity} and \ref{Fig:Pro_Q_cavity},  is computed using Eq.~\eqref{eq.Q_cavity}. Although the value of $Q$ is significantly larger than that in the micro-stripe geometry studied in the main text, the temperature and disorder dependence are very similar. This is expected because the dependence on temperature and disorder is largely controlled only by $1/\sigma_1$ \cite{devisser2014} and $Q \propto 1/\sigma_1$ in both cases. Therefore, based on those numerical results, we expect that the main findings of the paper apply to the experiments on both planar resonators and cavities, though the numerical values of $Q$ will be different.
	
	\section{The estimation of critical disorder by percolation and the energy gap} \label{sec:percolation}
	In this section, we estimate the critical disorder $V =V_c$ at which the transition occurs in order to determine whether the disorder strength $V\simeq 1.875$ for an optimal operation of the superinducting device is still on the metallic side of the transition. 
	For that purpose, we carry out a percolation analysis ~\cite{stauffer2003introduction} based on the site-dependent amplitude of the superconducting order parameter $\Delta(r)$. We compute the maximum disorder strength $V$ for which a percolating cluster exist.
	
\begin{figure}[!htbp]
		\begin{center}
			\subfigure[]{\label{fig.perco} 
				\includegraphics[width=7cm]{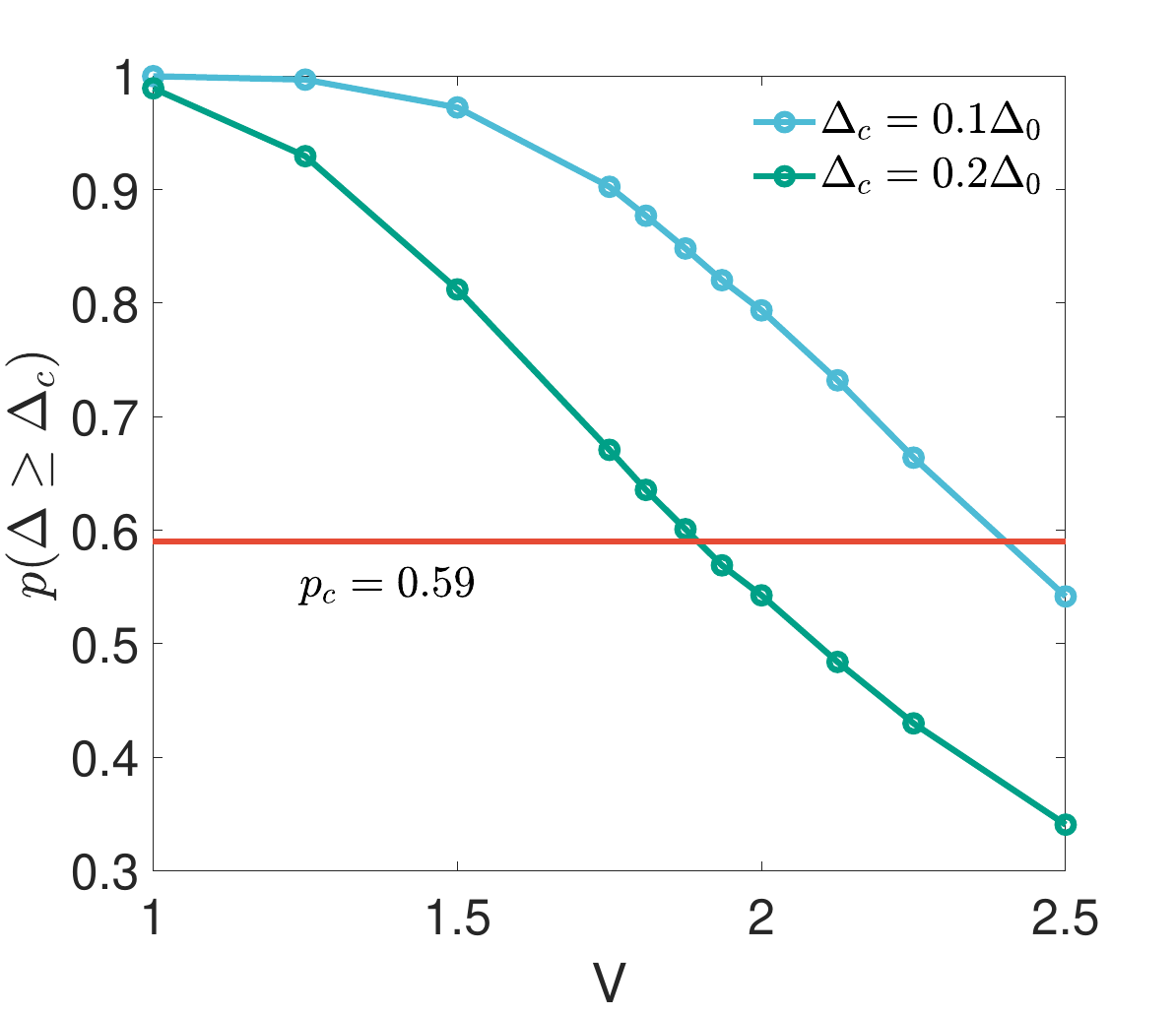}}
			\subfigure[]{\label{fig.gap} 
				\includegraphics[width=7cm]{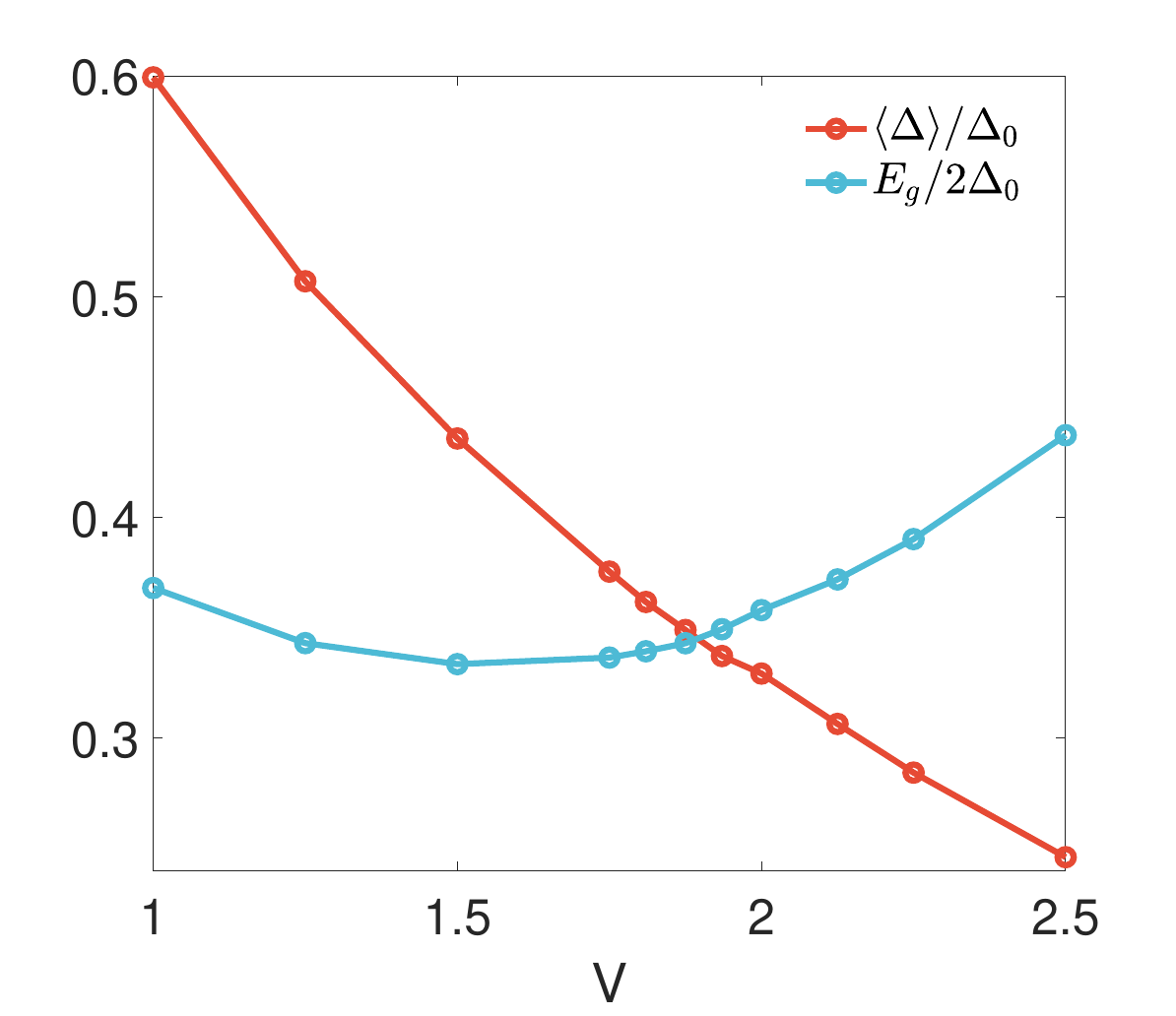}}
			\caption{\subref{fig.perco}. The probability of $\Delta(r) \ge \Delta_c$. In order to estimate the location of the transition, we show results for two cut-off values $\Delta_c = 0.1\Delta_0$ and $0.2\Delta_0$. The red line is the percolation threshold $p_c = 0.59$ for a simple 2D square lattice \cite{dean1967monte}. Based on the percolation results, the transition disorder is around $V=2$. \subref{fig.gap}. The spatial average of the order parameter $\langle \Delta \rangle $ (normalized by $\Delta_0$) and mean value of the spectral gap $E_g$ (normalized by $2\Delta_0$). As expected, the spatial averaged order parameter $\langle \Delta \rangle/ \Delta_0 $ decreases monotonously with disorder, and the spectral gap $E_g/2\Delta_0$ decreases with disorder first and then increases around $V=1.75$. The transition is expected to occur at this minimum of the spectral gap. To reduce statistical error, we  calculate 2000 disorder realizations for each $V$.}\label{Fig:perco_gap}
		\end{center}
	\end{figure}

	Strictly speaking, this is controlled by the probability $p$ that a given site has a finite value. 
	In a simple 2D lattice, the percolating transition occurs for $p_c=0.59$~\cite{dean1967monte}. In our case, we get finite values of $\Delta(i)$ for every site. However, we expect that if the order parameter is very small with respect to the clean value, it will not be able to carry a supercurrent because phase fluctuation will prevent local phase coherence. For that reason, we perform the percolation analysis with a cut-off value $\Delta_c$ of the order parameter, namely, we consider that a site belongs to the cluster only if $\Delta > \Delta_c$. Results depicted in Fig.~\ref{fig.perco} predict a critical disorder, which is relatively sensitive to the cutoff,  of about $V \simeq 2$.
	Another way to estimate the location of the transition is by computing the spectral gap as a function of the disorder strength. It is expected that it will decrease with disorder for weak disorder but this trend is reversed around the transition due to Anderson localization effects ~\cite{Ghosal2001}. The estimation of the critical disorder, $V=1.75$, by this method, see Fig.~\ref{fig.gap}, is consistent with the percolation calculation. Therefore, given the uncertainties of this estimates, the optimal disorder $V \simeq 1.875$ for the operation of the superinductor device seem to fall on the critical region but clearly not on the insulating side.

\end{document}